\documentstyle[seceq,psfig,wrapfig,axodraw]{ptptex}

\makeatletter
\def\maketitle{\par
 \begingroup
 \def\thefootnote{\fnsymbol{footnote}}
 \if@twocolumn
 \twocolumn[\@maketitle]
 \else \newpage
 \global\@topnum\z@ \@maketitle \fi\thispagestyle{plain}\@thanks
 \endgroup
 \let\maketitle\relax
 \let\@maketitle\relax
 \gdef\@thanks{}\gdef\@author{}\gdef\@title{}\let\thanks\relax}
\makeatother

\makeatletter

\newtoks\@stequation

\def\subequations{\refstepcounter{equation}%
  \edef\@savedequation{\the\c@equation}%
  \@stequation=\expandafter{\theequation}
  \edef\@savedtheequation{\the\@stequation}
  \edef\oldtheequation{\theequation}%
  \setcounter{equation}{0}%
  \def\theequation{\oldtheequation\alph{equation}}}

\def\endsubequations{%
  \ifnum\c@equation < 2 \@warning{Only \the\c@equation\space subequation
    used in equation \@savedequation}\fi
  \setcounter{equation}{\@savedequation}%
  \@stequation=\expandafter{\@savedtheequation}%
  \edef\theequation{\the\@stequation}%
  \global\@ignoretrue}

\def\eqnarray{\stepcounter{equation}\let\@currentlabel\theequation
\global\@eqnswtrue\m@th
\global\@eqcnt\z@\tabskip\@centering\let\\\@eqncr
$$\halign to\displaywidth\bgroup\@eqnsel\hskip\@centering
     $\displaystyle\tabskip\z@{##}$&\global\@eqcnt\@ne
      \hfil$\;{##}\;$\hfil
     &\global\@eqcnt\tw@ $\displaystyle\tabskip\z@{##}$\hfil
   \tabskip\@centering&\llap{##}\tabskip\z@\cr}

\def\dlinepattern#1#2{%
\ifdim#2<#1
   \errmessage{the 1st argument is less than the 2nd argument.}%
\else
   \gdef\dline@solid{#1}\gdef\dline@period{#2}%
\fi}

\def\dline#1{\@dline[#1]}
\def\@dline[#1-#2]{\noalign{\global\@dla#1\relax
\global\advance\@dla\m@ne
\ifnum\@dla>\z@\global\let\@gtempa\@dlinea\else
  \global\let\@gtempa\@dlineb\fi
\global\@dlb#2\relax
\global\advance\@dlb-\@dla}\@gtempa
\noalign{\vskip-\arrayrulewidth}}

\def\@dlinea{\multispan\@dla&\multispan\@dlb
\unskip\cleaders\hbox to \dline@period
{\hss\rule{\dline@solid}{\arrayrulewidth}\hss}\hfill\cr}

\newcount\@dla
\newcount\@dlb

\def\@dlineb{\multispan\@dlb
\unskip\cleaders\hbox to \dline@period
{\hss\rule{\dline@solid}{\arrayrulewidth}\hss}\hfill\cr}

\dlinepattern{2pt}{5pt}

\makeatother

\newcommand\textfrac[2]{{\textstyle\frac{#1}{#2}}}

\def\simgt{\rlap{\lower 3.5 pt \hbox{$\mathchar \sim$}}%
           \raise 1pt \hbox {$>$}}
\def\simlt{\rlap{\lower 3.5 pt \hbox{$\mathchar \sim$}}%
           \raise 1pt \hbox {$<$}}

\def\gev{{\,\rm GeV}}
\def\tev{{\,\rm TeV}}

\def\msbar{\overline{\rm MS }}	
\def\to{\rightarrow}

\def\mz{m_Z^{}}
\def\mw{m_W^{}}
\def\mh{m_H^{}}
\def\mmv{m_V^2 }
\def\mmz{m_Z^2 }
\def\mmw{m_W^2 }
\def\mmh{m_H^2 }
\def\ebar{\bar{e}}
\def\sbar{\bar{s}}
\def\cbar{\bar{c}}
\def\gzbar{\bar{g}_Z}
\def\gwbar{\bar{g}_W}
\def\ehat{\hat{e}}
\def\shat{\hat{s}}
\def\chat{\hat{c}}
\def\gzhat{\hat{g}_Z}
\def\ghat{\hat{g}}

\def\pibar{\overline{\Pi}}

\def\delb{\bar{\delta}_{b}}
\def\delg{\bar{\delta}_{G}^{}}
\def\zbb{Zb_L^{}b_L^{}}
\newcommand {\bea}	{\begin{eqnarray}}
\newcommand {\eea}	{\end{eqnarray}}
\newcommand {\be}	{\begin{equation}}
\newcommand {\ee}	{\end{equation}}
\newcommand {\bsub}	{\begin{subequations}}
\newcommand {\esub}	{\end{subequations}}
\def \ibid {{\it ibid}.}
\def \etal {{\it et al}.}

\def\gzbarrunning{%
  \begin{equation}
     {1}/{\gzbar^2(\mmz)} \approx {1}/{\gzbar^2(0)}
       -0.02390  +({2.41}/{m_t^2}) +({1.73}/{\mmh})
   \label{gzbarrunning}
  \end{equation}
}

\def\fitofgzbsb{%
  \begin{subequations}
   \label{fitofgzbsb}
  \begin{eqnarray}
     & &
     \left.
     \begin{array}{ll}
     \gzbar^2(\mmz) &\!\!= 0.55556
      -0.00049\,\textfrac{\alpha_s' -0.1081}{0.0043}
       \pm 0.00072
      \\[1mm]
     \sbar^2(\mmz)  &\!\!= 0.23041
      +0.00004\,\textfrac{\alpha_s' -0.1081}{0.0043}
       \pm 0.00029
     \end{array}
    \right\}\,\,
   \rho_{\rm corr} = 0.23,
   \\
   & & \quad
   \chi^2_{\rm min} =  15.6
         +\biggl(\frac{\alpha_s' -0.1081}{0.0043}\biggr)^2
         +\biggl(\frac{\delb-0.0025}{0.0042}\biggr)^2\,,
   \label{fitofgzbsbchisq}
  \end{eqnarray}
  \end{subequations}
}

\def\fitofmw{%
  \begin{equation}
     \label{fitofmw}
	\gwbar^2(0) = 0.4227\pm 0.0017\,,
  \end{equation}
}%

\def\fitofstu{%
  \begin{subequations}
     \label{fitofstu}
  \begin{eqnarray}
  && \left.
    \begin{array}{c@{}r@{}r@{}r@{}r}
\!\!   S =&   -0.42 &
              -0.059\,\frac{\alpha_s'-0.1093}{0.0042} &
              +0.06\,\frac{\delta_\alpha-0.03}{0.09} &
              \pm 0.15
         \\[1mm]
\!\!   T =&    0.57 &
              -0.104\,\frac{\alpha_s'-0.1093}{0.0042} & &
              \pm 0.17
         \\[1mm]
\!\!   U =&    0.16 &
              +0.079\,\frac{\alpha_s'-0.1093}{0.0042} &
              +0.02\,\frac{\delta_\alpha-0.03}{0.09} &
              \pm 0.49 \\[1mm]
    \end{array}
    \right\} \;
   \rho_{\rm corr} =\! \left(
      \begin{array}{rrr}
           1    &  0.86 & -0.10  \\[1mm]
                &  1    & -0.20  \\[1mm]
                &       &  1     \\[1mm]
      \end{array}
      \right), \,\,\,
   \nonumber \\ && \hspace{-5mm}
   \qquad \label{fitofstu_mean}
   \\
   & &\,\,\,\,\, \chi^2_{\rm min} =  20.4
         +\biggl(\frac{\alpha_s' -0.1093}{0.0042}\biggr)^2
         +\biggl(\frac{\delb-0.0025}{0.0042}\biggr)^2\,.
     \label{chisqofstu}
  \end{eqnarray}
  \end{subequations}
}%
\def\dataofnqccfrorig{%
\begin{equation}
   K = 0.5626 \pm 0.0025\,\mbox{(stat)} \pm 0.0036\,\mbox{(sys)}
              \pm 0.0028\,\mbox{(model)}\pm 0.0029\,\mbox{($m_c$)}.
   \label{dataofnqccfrorig}
\end{equation}
}

\def\fitoflenc{%
  \begin{subequations}
   \label{fitoflenc}
  \begin{eqnarray}
     & &
     \left.
     \begin{array}{ll}
     \gzbar^2(0) &= 0.5441 \pm 0.0029
      \\[1mm]
     \sbar^2(0)  &= 0.2362 \pm 0.0044
     \end{array}
    \right\}\quad
   \rho_{\rm corr} = 0.70,
   \\
   & & \quad
   \chi^2_{\rm min} =  2.7\,.
   \label{fitoflencchisq}
  \end{eqnarray}
  \end{subequations}
}

\def\fitoflencatmz{%
  \begin{subequations}
   \label{fitoflencatmz}
  \begin{eqnarray}
     & &
     \left.
     \begin{array}{ll}
     \gzbar^2(\mmz) &= 0.5512 \pm 0.0030
      \\[1mm]
     \sbar^2(\mmz)  &= 0.2277 \pm 0.0047
     \end{array}
    \right\}\quad
   \rho_{\rm corr} = 0.70,
   \\
   & & \quad
   \chi^2_{\rm min} =  2.7\,.
   \label{fitoflencatmzchisq}
  \end{eqnarray}
  \end{subequations}
}

\notypesetlogo  

\title{%
Looking Beyond the Standard Model\\ through\\ Precision Electroweak Physics%
}

\author{%
Kaoru Hagiwara
}

\inst{%
Theory Group, KEK, 1-1 Oho, Tsukuba, Ibaraki 305, Japan
}

\abst{%
The most important hint of physics beyond the Standard Model (SM)
from the 1995 precision electroweak data is that the most precisely
measured quantities, the total, leptonic and hadronic decay widths
of the $Z$ and the effective weak mixing angle, $\sin^2\theta_W$,
measured at LEP and SLC, and the quark-lepton universality of
the weak charged currents measured at low energies, all agree
with the predictions of the SM at a few $\times 10^{-3}$ level.
By taking into account the above constraints I examine implications
of three possible disagreements between experiments and the SM
predictions.
It is difficult to interpret the 11\% (2.5-$\sigma$) deficit of
the $Z$-partial-width ratio $R_c=\Gamma_c/\Gamma_h$, since it
either implies an unacceptably large $\alpha_s$ or a subtle
cancellation among hadronic $Z$ decay widths in order to keep
all the other successful predictions of the SM.
The 2\% (3-$\sigma$) excess of the ratio $R_b=\Gamma_b/\Gamma_h$
may indicate the presence of a new rather strong interaction,
such as the top-quark Yukawa coupling in the supersymmetric
(SUSY) SM or a new interaction responsible for the large
top-quark mass in the Technicolor scenario of dynamical
electroweak symmetry breaking.
Another interpretation may be additional tree-level gauge
interactions that couple only to the third generation of fermions.
A common consequence of these attempts is a rather small $\alpha_s$,
$\alpha_s(\mz )_{\msbar}=0.104\pm 0.08$.
The 0.17\% (1-$\sigma$) deficit of the CKM unitarity relation
$|V_{ud}|^2+|V_{us}|^2+|V_{ub}|^2=1$ may indicate light
sleptons and gauginos in the minimal SUSY-SM.
None of the above disagreements are, however, convincing
at present.
}

\begin{document}

\thispagestyle{empty}
\vspace*{-2cm}
\begin{flushright}
\begin{tabular}{l}
KEK--TH--463         \\[-1mm]
KEK preprint 95--186 \\[-1mm]
January 1996         \\[-1mm]
H
\end{tabular}
\end{flushright}
\vspace*{15mm}

\begin{center} \Large \bf
Looking Beyond the Standard Model  \\
through                            \\
Precision Electroweak Physics
\end{center}

\vspace*{8mm}

\begin{center}
\large \bf Kaoru Hagiwara
\end{center}
\vspace*{2mm}
\begin{center}
{\it Theory Group, KEK, 1-1 Oho, Tsukuba, Ibaraki 305, Japan }
\end{center}

\vspace*{15mm}
\begin{center}
\bf ABSTRACT
\end{center}
\vspace*{5mm}

\baselineskip=13pt
\begin{center}
\begin{minipage}{13cm}
The most important hint of physics beyond the Standard Model (SM)
from the 1995 precision electroweak data is that the most precisely
measured quantities, the total, leptonic and hadronic decay widths
of the $Z$ and the effective weak mixing angle, $\sin^2\theta_W$,
measured at LEP and SLC, and the quark-lepton universality of
the weak charged currents measured at low energies, all agree
with the predictions of the SM at a few $\times 10^{-3}$ level.
By taking into account the above constraints I examine implications
of three possible disagreements between experiments and the SM
predictions.
It is difficult to interpret the 11\% (2.5-$\sigma$) deficit of
the $Z$-partial-width ratio $R_c=\Gamma_c/\Gamma_h$, since it
either implies an unacceptably large $\alpha_s$ or a subtle
cancellation among hadronic $Z$ decay widths in order to keep
all the other successful predictions of the SM.
The 2\% (3-$\sigma$) excess of the ratio $R_b=\Gamma_b/\Gamma_h$
may indicate the presence of a new rather strong interaction,
such as the top-quark Yukawa coupling in the supersymmetric
(SUSY) SM or a new interaction responsible for the large
top-quark mass in the Technicolor scenario of dynamical
electroweak symmetry breaking.
Another interpretation may be additional tree-level gauge
interactions that couple only to the third generation of fermions.
A common consequence of these attempts is a rather small $\alpha_s$,
$\alpha_s(\mz )_{\msbar}=0.104\pm 0.08$.
The 0.17\% (1-$\sigma$) deficit of the CKM unitarity relation
$|V_{ud}|^2+|V_{us}|^2+|V_{ub}|^2=1$ may indicate light
sleptons and gauginos in the minimal SUSY-SM.
None of the above disagreements are, however, convincing
at present.
\end{minipage}
\end{center}

\vspace*{30mm}
\begin{center}
{\it Talk presented at Yukawa International Seminar (YKIS)~'95 \\
``From the Standard Model to Grand Unified Theories'' \\
Kyoto, Japan, August 21--25, 1995 }
\end{center}
\setcounter{page}{0}

\maketitle

\section{Introduction}

Despite the firm theoretical belief that the Standard Model (SM)
of the electroweak interactions is merely an effective low-energy
description of a more fundamental theory,
high-energy experiments have so far been unable to establish
a signal of new physics.
Naturalness of the dynamics of the electroweak-gauge-symmetry
breakdown suggests that the energy scale of new-physics
should lie below or at $\sim 1$~TeV.
Because of this relatively low new-physics scale,
there has been a hope that hints of new physics beyond
the SM might be found as quantum effects affecting precision
electroweak observables.

In response to such general expectations the experimental
accuracy of the electroweak measurements has steadily been
improved in the past several years,
reaching the $10^{-5}$ level for $\mz$,
a few $\times 10^{-3}$ level for the total and
some of the partial $Z$ widths,
and the $10^{-2}$ level for the asymmetries at LEP and SLC.
Because of partial cancellation in the observable asymmetries
at LEP and SLC, their measurements at the $10^{-2}$ level
determine the effective electroweak mixing parameter,
$\sin^2\theta_W$, at the $10^{-3}$ level.
Therefore, by choosing the fine structure constant, $\alpha$,
the muon-decay constant, $G_F$, and $\mz$ as the three inputs
whose measurement error is negligibly small,
we can test the predictions of the SM at
a few $\times 10^{-3}$ level.
Accuracy of experiments has now reached the level where
new physics contributions to quantum corrections
can be probed.

In this report I would like to summarize the electroweak
measurements that were reported as preliminary results for the
1995 Summer Conferences\cite{lepewwg9502,lephf9502,lp95_renton}
from a theorist's point of view\cite{lp95_hag}
and to examine their implications on our search for new physics.
In Section~2 we summarize the latest electroweak results
at LEP, SLC, the Tevatron, and at low energies.
These data are analysed in the general
${\rm SU(2)_L\times U(1)_Y}$ model framework\cite{hhkm}.
In Section~3 we discuss the nature of the $R_b$ and $R_c$
`crisis' in detail.
In Section~4 I introduce several theoretical attempts
to explain the observed 2\% (3-$\sigma$) excess in $R_b$.
In Section~5 I examine implications of a possible (1-$\sigma$)
violation of the quark-lepton universality in the low-energy
charged-current experiments.
Section~6 summarises our findings.



\section{ The 1995 Precision Electroweak Data }
%
In this section we analyse the preliminary electroweak results%
\cite{lepewwg9502,lephf9502,lp95_renton} presented at the 1995
summer conferences in the general ${\rm SU(2)_L\times U(1)_Y}$
model framework\cite{hhkm}.
We allow a new physics contribution to the $S$, $T$, $U$
parameters\cite{stu} of the electroweak gauge-boson-propagator
corrections as well as to the $Zb_Lb_L$ vertex form factor,
$\delb(\mmz )$, but otherwise we assume the SM contributions
dominate the corrections.
We take the strengths of the QCD and QED couplings at the $\mz$
scale, $\alpha_s(\mz )$ and $\bar{\alpha}(\mmz )$, as external
parameters of the fits so that implications of their
precise measurements on electroweak physics are manifestly shown.
Those who are familiar with the framework of Ref.\citen{hhkm}
may skip the following subsection.

\subsection{ Brief Review of Electroweak Radiative Corrections
	     in $\rm{SU(2)_L \times U(1)_Y}$ Models }

The propagator corrections in the general
$\rm{SU(2)_L \times U(1)_Y}$
models can conveniently be expressed in terms of the following
four effective charge form-factors\cite{hhkm}:
\vspace*{-8mm}
\def\propagator#1#2{%
    \begin{picture}(80,35)(0,18)
       \Text(20,27)[cb]{#1}
       \Text(50,27)[cb]{#2}
       \Line(0,35)(10,20)
       \Line(0, 5)(10,20)
       \Line(70,35)(60,20)
       \Line(70, 5)(60,20)
       \Photon(10,20)(30,20){3}{3}
       \Photon(40,20)(60,20){3}{3}
       \GCirc(35,20){7}{0.5}
    \end{picture}
}
\bsub \label{barcharges}
  \begin{eqnarray}
    \propagator{$\gamma$}{$\gamma$} &\sim& \ebar^2(q^2) =\ehat^2
    \Bigl[\,1-{\rm Re}\pibar_{T,\gamma}^{\gamma\gamma}(q^2)\,\Bigr]\,,
\\[-2mm]
    \propagator{$\gamma$}{$Z$}      &\sim& \sbar^2(q^2)      =\shat^2
    \Bigl[\,1+\shat\chat\,{\rm Re}\,\pibar_{T,\gamma}^{\gamma Z}(q^2)
    \,\Bigr]\,
\\[-2mm]
    \propagator{$Z$}{$Z$}           &\sim& \gzbar^2(q^2)     =\gzhat^2
    \Bigl[\,1 -{\rm Re}\pibar_{T,Z}^{ZZ}(q^2)\,\Bigr]\,,
\\[-2mm]
    \propagator{$W$}{$W$}           &\sim& \gwbar^2(q^2)     =\ghat^2
    \Bigl[\,1 -{\rm Re}\pibar_{T,W}^{WW}(q^2)\,\Bigr]\,,
  \end{eqnarray}
\esub
where
$
     \pibar_{T,V}^{AB}(q^2)
      \equiv [\pibar_T^{AB}(q^2)\!-\!\pibar_T^{AB}(\mmv)]/(q^2\!-\!\mmv)
$
are the propagator correction factors that appear in the
$S$-matrix elements after the mass renormalization is performed,
and $\ehat \equiv \ghat\shat \equiv \gzhat^{}\shat\chat$
are the $\msbar$ couplings.
The `overlines' denote the inclusion of the pinch terms%
\cite{pinch,pinch2,kl89}, which make these effective charges
useful\cite{kl89,hhkm,hms95} even at very high energies
($|q^2|\gg \mmz$).
The amplitudes are then expressed in terms of these charge
form-factors plus appropriate vertex and box corrections.
In our analysis\cite{hhkm} we assume that all the vertex
and box corrections are dominated by the SM contribution
except for the $\zbb$ vertex,
\bea \label{zblbl}
\Gamma_L^{Zbb}(q^2) = -\gzhat \{ -\frac{1}{2}[1+\delb(q^2)]
		+\frac{1}{3}\shat^2[1+\Gamma_1^{b_L}(q^2)] \} ,
\eea
for which the function  $\delb(\mmz )$ is allowed to take
an arbitrary value.
The charge form-factors and $\delb(\mmz )$ can then be extracted
from the experimental data, and the extracted values can
be compared with various theoretical predictions.

We can {\rm define}\cite{hhkm} the $S$, $T$ and $U$ variables
in terms these effective charges,
  \begin{subequations}\label{stu_def}
  \begin{eqnarray}
    \frac{\sbar^2(\mmz)\cbar^2(\mmz)}{\bar{\alpha}(\mmz)}
    -\frac{4\,\pi}{\gzbar^2(0)} &\equiv& \;\frac{S}{4} \,,
    \\
    \frac{\sbar^2(\mmz)}{\bar{\alpha}(\mmz)}\quad
    -\frac{4\,\pi}{\gwbar^2(0)} &\equiv& \frac{S+U}{4} \,,
    \\
    1\;-\,\frac{\gwbar^2(0)}{\mmw}\frac{\mmz}{\gzbar^2(0)}
    &\equiv& \;\alpha T \,,
  \end{eqnarray}
  \end{subequations}
where $\bar{\alpha}(q^2)\!\equiv\!\ebar^2(q^2)/4\pi$, and
it is clear that these variables measure deviations
from the naive universality of the electroweak gauge couplings.
They receive contributions from both the SM radiative effects
as well as new physics contributions.
The original $S$, $T$, $U$ variables of Ref.\citen{stu} are
obtained in Ref.\citen{hhkm} approximately by subtracting
the SM contributions (at $\mh=1000$~GeV).

For a given electroweak model we can calculate the $S$, $T$,
$U$ parameters ($T$ is a free parameter in models without
custodial SU(2) symmetry), and the charge form-factors
are then fixed by the following identities\cite{hhkm}:
  \begin{subequations}
    \label{gbarfromstu}
  \begin{eqnarray}
     \frac{1}{\gzbar^2(0)}
        &=& \frac{1+\delg -\alpha \,T}{4\,\sqrt{2}\,G_F\,\mmz} \,,
    \label{gzbarfromt}\\[2mm]
      \sbar^2(\mmz)
          &=& \frac{1}{2}
              -\sqrt{\frac{1}{4} -\bar{\alpha}^2(\mmz)
                    \biggl(\frac{4\,\pi}{\gzbar^2(0)} +\frac{S}{4}
                    \biggr)  }\,,
    \label{sbarfroms}\\[2mm]
       \frac{4\,\pi}{\gwbar^2(0)}
             &=& \frac{\sbar^2(\mmz)}{\bar{\alpha}^2(\mmz)}
                -\frac{1}{4}\,(S+U) \,.
    \label{gwbarfromu}
  \end{eqnarray}
  \end{subequations}
Here $\delg$ ($\delg=0.0055$ in the SM) is the vertex and box correction
to the muon lifetime\cite{del_gf} after subtraction of the pinch
term\cite{hhkm}:
 \begin{eqnarray}
   G_F &=& \frac{\gwbar^2(0)+\ghat^2\delg}{4\,\sqrt{2}\mmw} \,.
   \label{gf}
 \end{eqnarray}

It is clear from the above identities that, once we know
$T$ and $\delg$ in a given model, we can predict
$\gzbar^2(0)$, and then, by knowing $S$ and $\bar{\alpha}(\mmz)$,
we can calculate $\sbar^2(\mmz)$, and finally, by knowing
$U$, we can calculate $\gwbar^2(0)$.
Since $\bar{\alpha}(0)=\alpha$ is known precisely, all
four charge form-factors are then fixed at one $q^2$ point.
The $q^2$-dependence of the form factors should also be
calculated in a given model, but it is less dependent
on physics at very high energies\cite{hhkm}.
In the following analysis we assume that the SM contribution
governs the running of the charge form-factors between
$q^2=0$ and $q^2=\mmz$\footnote{%
Analyses which allow new physics contributions to the running
of the charge form factors are found e.g.\ in
Refs.\citen{hisz93,hms95}.
}.  We can now predict all the neutral-current amplitudes
in terms of $S$ and $T$, and an additional knowledge of
$U$ gives the $W$ mass via Eq.(\ref{gf}).

We should note here that our prediction for the effective
mixing parameter, $\sbar^2(\mmz)$, is not only sensitive to
the $S$ and $T$ parameters but also to the precise value
of $\bar{\alpha}(\mmz)$.
This is the reason why the predictions for the asymmetries
measured at LEP/SLC and, consequently, the experimental
constraint on $S$ extracted from the asymmetry data
are dependent on $\bar{\alpha}(\mmz)$.
In order to parametrize the uncertainty in our evaluation
of $\bar{\alpha}(\mmz)$, the parameter $\delta_\alpha$
is introduced in Ref.\citen{hhkm} as follows:
\bea
\label{del_a}
\delta_\alpha \equiv 1/\bar{\alpha}(\mmz) -128.72 \,.
\eea
We show in Table~\ref{tab:alpha_mz} the results of the
four recent updates\cite{mz94,swartz95,eidjeg95,bp95}
on the hadronic contribution to the running of the effective
QED coupling.
Three definitions of the running QED coupling are compared.
I remark that our simple formulae
(\ref{stu_def}) and (\ref{gbarfromstu}) are valid only if one
includes all the fermionic and bosonic contributions
to the propagator corrections.
There is no discrepancy among the four recent estimates
in Table~\ref{tab:alpha_mz}, where small differences are attributed
to the use of perturbative QCD for constraining the magnitude
of medium energy data\cite{mz94} or to a slightly different set
of input data\cite{swartz95}.
For more detailed discussions I refer the readers to
an excellent review by Takeuchi\cite{takeuchi95}.
In the following analysis we take the estimate of
Ref.\citen{eidjeg95} ($\delta_\alpha=0.03\pm 0.09$)
as the standard, and we show the sensitivity of our results
to $\delta_\alpha-0.03$.

\begin{table}[t]
\caption{%
The running QED coupling at the $\mz$ scale in the three schemes.
$\alpha(\mmz)_{\rm l.f.}$ contains only the light fermion
contributions to the running of the QED coupling constant
between $q^2\!=\!0$ and $q^2\!=\!\mmz$.
$\alpha(\mmz)_{\rm f}$ contains all fermion contributions
including the top-quark.
$m_t\!=\!175$~GeV and $\alpha_s(\mz)\!=\!0.12$ are assumed
in the perturbative two-loop correction\protect\cite{kniehl90}.
$\bar{\alpha}(\mmz)$ contains also the $W$-boson-loop
contribution\protect\cite{hhkm}
including the pinch term\protect\cite{pinch,pinch2}.
$\delta_\alpha\!\equiv\!1/\bar{\alpha}(\mmz )\!-\!128.72$.
}
\label{tab:alpha_mz}
\begin{center}
{\footnotesize
\begin{tabular}{|c|c|c|c|c|}
\hline
& $1/\alpha(\mmz)_{\rm l.f.}$
& $1/\alpha(\mmz)_{\rm f}$
& $1/\bar{\alpha}(\mmz)$
& $\delta_\alpha$
\\
\hline
Martin-Zeppenfeld '94\cite{mz94}
& $128.98\pm 0.06$ & $128.99\pm 0.06$ & $128.84\pm 0.06$ & $0.12\pm 0.06$
\\
Swartz '95\cite{swartz95}
& $128.96\pm 0.06$ & $128.97\pm 0.06$ & $128.82\pm 0.06$ & $0.10\pm 0.06$
\\
Eidelman-Jegerlehner '95\cite{eidjeg95}
& $128.89\pm 0.09$ & $128.90\pm 0.09$ & $128.75\pm 0.09$ & $0.03\pm 0.09$
\\
Burkhardt-Pietrzyk '95\cite{bp95}
& $128.89\pm 0.10$ & $128.90\pm 0.10$ & $128.76\pm 0.10$ & $0.04\pm 0.10$
\\
\hline
\end{tabular}
}
\end{center}
\end{table}

Once we know $\bar{\alpha}(\mmz)$ the charge form-factors
in Eq.(\ref{gbarfromstu}) can be calculated from $S$, $T$, $U$.
The following approximate formulae\cite{hhkm} are useful:
  \begin{subequations}
   \label{gbar_approx}
  \begin{eqnarray}
        \gzbar^2(0)    &\approx& 0.5456 \hphantom{+0.0036\,S}\;\,
            +0.0040\,T'\,,
   \label{gzbar_approx}\\
        \sbar^2(\mmz) &\approx& 0.2334            +0.0036\,S
            -0.0024\,T'
           \hphantom{ +0.0035\,U }\;\, -0.0026\,\delta_\alpha    \,,
         \qquad
   \label{sbar_approx}\\
        \gwbar^2(0)    &\approx& 0.4183            -0.0030\,S
            +0.0044\,T'
            +0.0035\,U               +0.0014\,\delta_\alpha      \,,
         \qquad
   \label{gwbar_approx}
  \end{eqnarray}
  \end{subequations}
where we note that the predictions depend on $T$ only through
the combination\cite{hhkm}
\bea \label{t'}
	T'\equiv T+(0.0055-\delg)/\alpha \,.
\eea
The values of $\gzbar^2(\mmz)$ and $\sbar^2(0)$ are then
calculated from $\gzbar^2(0)$ and $\sbar^2(\mmz)$ above,
respectively, by assuming the SM running of the form-factors.
The $Z$ widths are sensitive to $\gzbar^2(\mmz)$, which can
be obtained from $\gzbar^2(0)$ in the SM approximately by
\gzbarrunning
when $m_t(\gev )>150$ and $\mh(\gev )>100$.
Details of the following analysis will be reported elsewhere\cite{hhm96}.

\subsection{ Key Observations from the 1995 Electroweak Data }
%
The 1995 update of the precision electroweak data are found in
the LEP Electroweak Working Group reports%
\cite{lepewwg9502,lephf9502}
in which preliminary results from LEP, SLC and
the Tevatron are combined.
Its concise summary is given in Ref.\citen{lp95_hag}.

I would like to summarize the results by the following four key observations:
\begin{itemize}
\item{
The three line-shape parameters, $\Gamma_Z$, $\sigma_h^0$
and $R_l$, are now measured with an accuracy better than 0.2\%.
\bsub \label{del_lineshape}
\bea
{\Delta\Gamma_Z}/{\Gamma_Z}     &=&          - 0.0010 \pm 0.0013 \,,\\
{\Delta\sigma_h^0}/{\sigma_h^0} &=& \phantom{-}0.0006 \pm 0.0019 \,,\\
{\Delta R_l}/{R_l}              &=& \phantom{-}0.0015 \pm 0.0015 \,,
\eea
\esub
where $\Delta$ gives the difference between the data and the reference
SM predictions\cite{hhkm} for $m_t=175$~GeV, $\mh=100$~GeV,
$\alpha_s=0.12$ and $\delta_\alpha=0.03$.
We should note here that the above three most accurately
measured line-shape parameters constrain the $Z$ partial
widths $\Gamma_l$, $\Gamma_h$ and $\Gamma_{\rm inv}$
accurately;
\bsub \label{del_zwidths}
\bea
\Delta \Gamma_h/\Gamma_h                 &=& \phantom{-}0.0001 \pm 0.0017 \,,\\
\Delta \Gamma_l/\Gamma_l                 &=&          - 0.0013 \pm 0.0016 \,,\\
\Delta \Gamma_{\rm inv}/\Gamma_{\rm inv} &=&          - 0.004  \pm 0.005  \,,
\label{del_inv}
\eea
\esub
because they are three independent combinations of the above
three widths,
$\Gamma_Z =\Gamma_h\!+\!\Gamma_l\!+\!\Gamma_{\rm inv}$,
$R_l=\Gamma_h/\Gamma_l$, and
$\sigma_h^0 = (12\pi/\mmz )\Gamma_h\Gamma_l/\Gamma_Z^2$.
}
\item{
Detailed tests\cite{lepewwg9502,lp95_renton} of the $e$-$\mu$-$\tau$
universality show that any hint of universality violation in the
1994 data is now disappearing.
\bsub \label{e_mu_tau}
\bea
{\Delta\Gamma_\tau}/{\Gamma_\tau}             &=&\phantom{-}0.000\pm 0.0035
\,,\\
{\Delta\Gamma_{\nu_\tau}}/{\Gamma_{\nu_\tau}} &=&         - 0.012\pm 0.015
\,,\\
{\Delta A_\tau}/{A_\tau}                      &=&         - 0.04 \pm 0.05
\,,\\
{\Delta A_{\rm FB}^\tau}/{A_{\rm FB}^\tau}    &=&\phantom{-}0.23 \pm 0.14   \,.
\eea
\esub
The result for $\Gamma_{\nu_\tau}$ is obtained from Eq.(\ref{del_inv})
by assuming
$\Gamma_{\rm inv}=\Gamma_{\nu_e}+\Gamma_{\nu_\mu}+\Gamma_{\nu_\tau}$
with the SM values for $\Gamma_{\nu_e}$ and $\Gamma_{\nu_\mu}$.
Although the $\tau$ forward-backward asymmetry is still 1.7$\sigma$
away from the SM prediction, the accuracy of the measurement is still
poor (14\%), and its significance is overshadowed by the excellent agreements
in the partial widths $\Gamma_\tau$, $\Gamma_{\nu_\tau}$ and
the $\tau$ polarization asymmetry, $A_\tau$, which are measured at
the 0.35\%, 1.5\% and 5\% level, respectively.
}
\begin{figure}[t]
\begin{center}
  \leavevmode\psfig{file=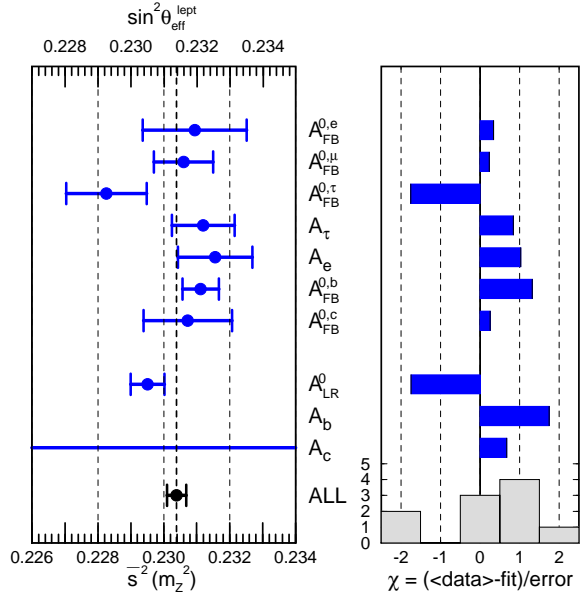,height=8cm,silent=0}
\vspace{3mm}
\caption{%
The effective electroweak mixing parameter\protect\cite{hhkm}
${\sbar}^2(\mmz)$ is determined from all the asymmetry
data from LEP and SLC.
The effective parameter $\sin^2\theta_{\rm eff}^{\rm lept}$
of the LEP Electroweak Working Group\protect\cite{lepewwg9502}
is obtained accurately\protect\cite{hhkm} as
$\sin^2\theta_{\rm eff}^{\rm lept} = {\sbar}^2(\mmz) +0.0010$.
The data on $A_b$ is off the scale.
}
\label{fig:sb}
\end{center}
\end{figure}
\item{
All the asymmetry data, including the left-right beam-polarization
asymmetry, $A_{\rm LR}$, from SLC, are now consistent with each other.
I show in Fig.~\ref{fig:sb} the result of the one-parameter fit to
all the asymmetry data in terms of the effective electroweak
mixing-angle, $\sbar^2(\mmz )$.
The fit gives
\bea
\label{sb2_fit95}
	\sbar^2(\mmz ) = 0.23039 \pm 0.00029
\eea
with $\chi^2_{\rm min}/({\rm d.o.f.})=13.0/(9)$.
The updated measurements of the asymmetries agree well (16\%CL)
with the ansatz that the asymmetries are determined by the
universal electroweak mixing parameter.
Even though $\sbar^2(\mmz)$ determined from the left-right
beam-polarization asymmetry of the $b$-quark forward-backward
asymmetry, $A_b$, is off the scale,
its significance is still moderate because of its large
measurement error, $\Delta A_b/A_b=-0.10\pm 0.06$.
}
\item{
The only data which disagree significantly with the predictions
of the SM are the two ratios $R_b=\Gamma_b/\Gamma_h$ and
$R_c=\Gamma_c/\Gamma_h$.
Here $\Gamma_q$ denotes the partial $Z$ decay widths into
$q\bar{q}$-initiated hadronic final states, and
$\Gamma_h$ denotes the hadronic $Z$ decay width.
$R_b$ is larger than the reference SM prediction
by 3\% (3.7-$\sigma$), whereas $R_c$ is smaller than
the prediction by 11\% (2.5-$\sigma$).
The trends of larger $R_b$ and smaller $R_c$ existed in the
combined data for the past few years, but their significance
grew considerably in the 1995 update.
}
\end{itemize}
I remark here that the interpretation of the possible deviations
from the SM in the last item is severely restricted by the
remaining excellent successes of the SM listed above.
I will show in Section~3 that it is difficult to accommodate
the $R_c$ data with all the other successes of the SM,
but the $R_b$ data can easily be accommodated by modifying
only the $\zbb$ vertex.

In this section I show results of the analysis\cite{hhm96}
where all the vertex and box corrections, except for the
$\zbb$ vertex function, $\delb(\mmz )$, are dominated by
the SM contributions.
All the LEP and SLC results are then fitted by the four
parameters
$\gzbar^2(\mmz )$, $\sbar^2(\mmz )$, $\delb(\mmz )$ and
$\alpha_s=\alpha_s(\mz )_{\msbar}$.
We find\cite{hhm96}
%
\fitofgzbsb
%
where
\bea
\label{alps'}
\alpha_s' = \alpha_s(\mz )_{\msbar}\, +1.54\,\delb(\mmz )
\eea
is the combination\cite{hhkm} that appears in the theoretical
prediction for $\Gamma_h$.
As a consequence of the $R_b$ data the best fit is obtained
at $\delb(\mmz )=0.0025$ and $\alpha_s=0.1043$.
Within the SM, however, the form factor $\delb(\mmz )$
takes only a negative value ($\delb(\mmz )\simlt -0.03$), and
its magnitude grows quadratically with $m_t$.
It can be parametrized accurately in the region
100$<m_t(\gev )<$200 as\cite{hhkm}
\bea
\label{delb_sm}
\delb(\mmz )_{\rm SM} \approx -0.00099-0.00211(\frac{m_t+31}{100})^2 \,.
\eea
We find e.g.\ $\alpha_s=0.1234\pm0.0043$ for $m_t=175$~GeV.

\begin{figure}[t]
\begin{minipage}[t]{7.2cm}
 \begin{center}
 \leavevmode\psfig{file=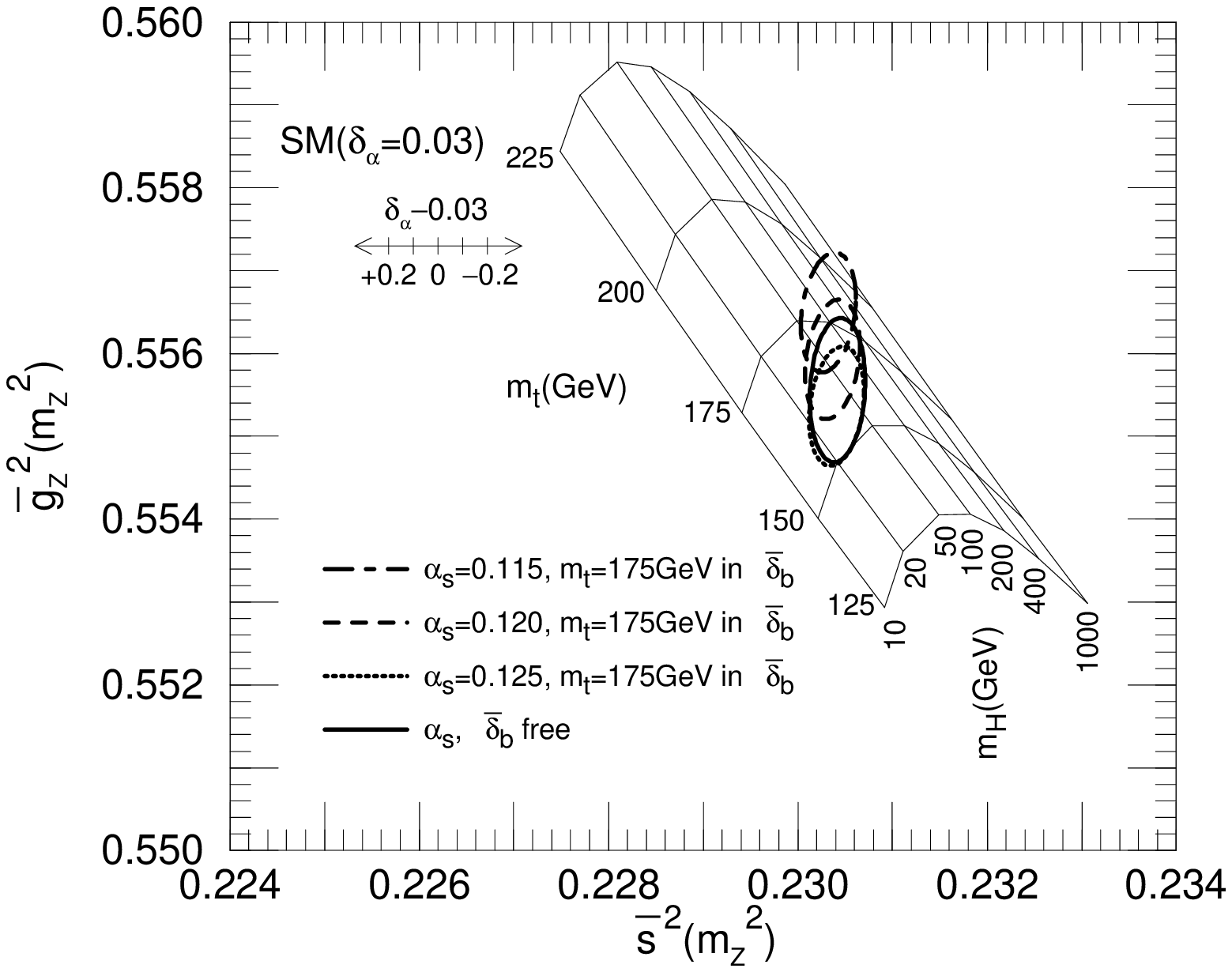,width=7.1cm,silent=0}
 \end{center}
\caption{%
A two-parameter fit to the $Z$ boson parameters in the
($\protect\sbar^2(\mmz), \protect\gzbar^2(\protect\mmz)$)
plane gives the 1-$\sigma$ contours for
$\alpha_s\!=\!0.115$, 0.120, 0.125 when
the $\protect\zbb$ vertex form-factor,
$\protect\delb(\protect\mmz)$, is evaluated in the SM
for $m_t\!=\!175$\,GeV.
The solid contour is obtained by a four-parameter fit where
both $\alpha_s$ and $\delb(\mmz )$ are allowed to vary.
The SM predictions at $\delta_\alpha \!=\!0.03$
and their dependences on $\delta_\alpha\!-\!0.03$
are also given.
}
\label{fig:gzbsb}
\end{minipage}
\hfill
\hspace*{1mm}
\begin{minipage}[t]{7.2cm}
 \begin{center}
 \leavevmode\psfig{file=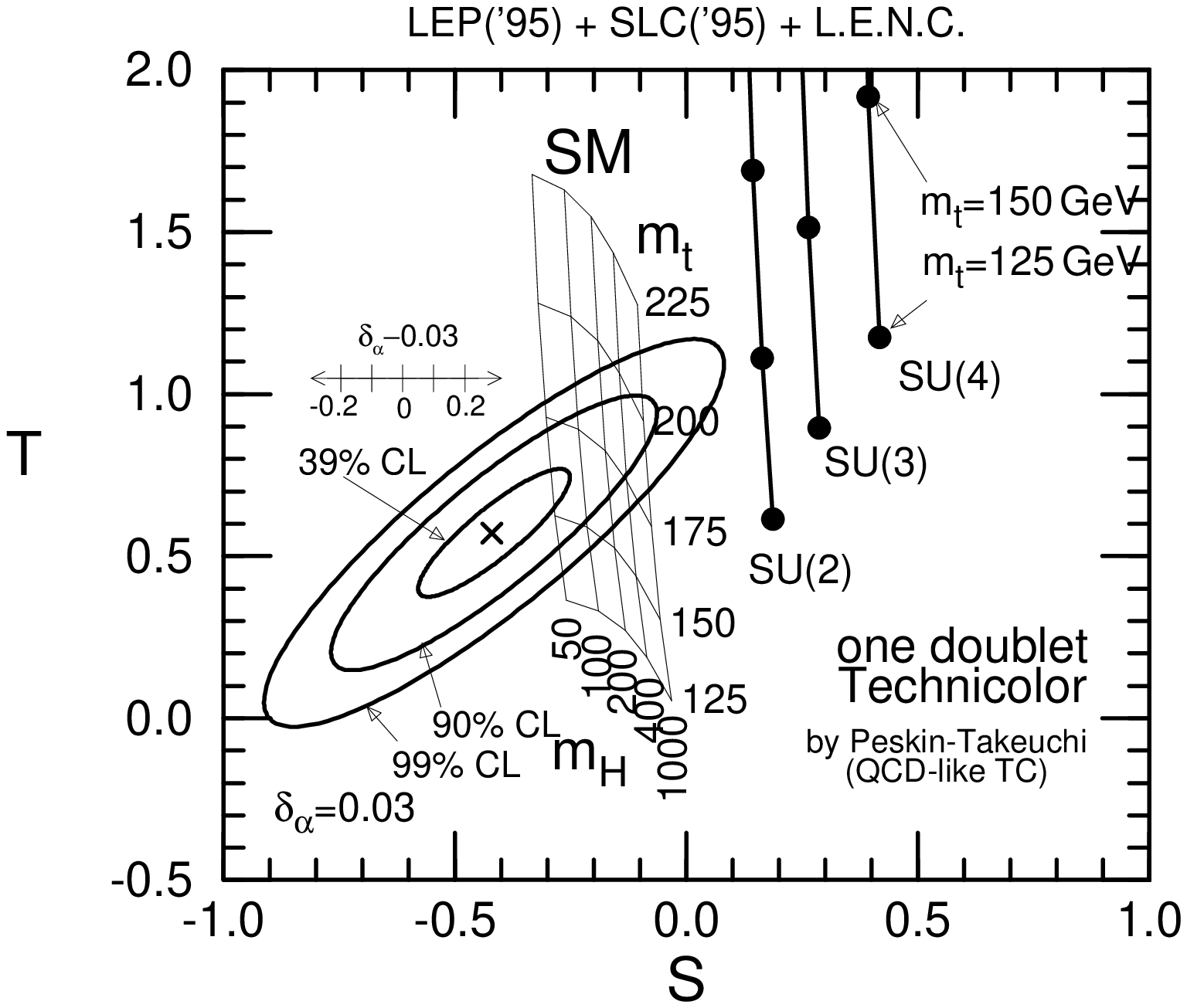,width=7.0cm,silent=0}
 \end{center}
\caption{%
Constraints on ($S$, $T$) from the five-parameter fit to
all the electroweak data for $\delta_\alpha=0.03$ and
$\protect\delg=0.0055$.
Together with $S$ and $T$ the $U$ parameter,
the $\protect\zbb$ vertex form-factor,
$\protect\delb(\protect\mmz)$, and the QCD coupling,
$\alpha_s(\mz)$, are allowed to vary in the fit.
Also shown are the SM predictions\protect\cite{hhkm}
and the predictions\protect\cite{stu} of one-doublet
${\protect\rm SU}(N_c)$--TC models for $N_c=2,3,4$.
}
\label{fig:st}
\end{minipage}
\end{figure}

In Fig.~\ref{fig:gzbsb} we show the 1-$\sigma$ (39\%CL) allowed
contours for $\alpha_s=0.115$, 0.120, 0.125 when $\delb$ takes its
SM value, $-\!0.010$, at $m_t\!=\!175$\,GeV.
If we allow {\em both} $\delb(\mmz )$ and $\alpha_s$ to be
freely fitted by the data, we obtain the thick solid contour.
The SM predictions for $\delta_\alpha=0.03$ and their
dependence on $\delta_\alpha\!-\!0.03$ are also given.
As expected, only $\sbar^2(\mmz )$ is sensitive to
$\delta_\alpha$.

The fit from the low-energy neutral-current data is
updated\cite{hhm96} by including the new CCFR data\cite{ccfr95_kevin}:
%
\fitoflenc
%
More discussions on the role of the low-energy neutral-current
experiments are given in the following subsection.

The $W$ mass data,
$\mw = 80.26 \pm 0.16 \gev$,
gives
%
\fitofmw
for $\delg=0.0055$ from Eq.(\ref{gf}).

We can now regard the fits
(\ref{fitofgzbsb}), (\ref{fitoflenc}), (\ref{fitofmw})
and the estimate\cite{eidjeg95} $\delta_\alpha=0.03\pm0.09$
for $\bar{\alpha}(\mmz)$ via Eq.(\ref{del_a})
as a parametrization of all the electroweak data in terms
of the four charge form-factors.
We then perform a five-parameter fit to all the electroweak data
in terms of $S$, $T$, $U$, $\delb$ and $\alpha_s$,
by assuming that the running of the charge form-factors is
governed by the SM contributions.
We find
%
\fitofstu
%
The dependence of the $S$ and $U$ parameters upon the external
parameter $\delta_\alpha$ of the fit may be understood from
Eq.(\ref{gbar_approx}).
For an arbitrary value of $\delg$ the parameter $T$ should be
replaced by $T'$ of Eq.(\ref{t'}).
It should be noted that the uncertainty in $S$
coming from $\delta_\alpha=0.03\pm 0.09$ is
of the same order as that from the uncertainty in $\alpha_s$;
they are not negligible when compared to the overall error.
The $T$ parameter has little $\delta_\alpha$ dependence,
but it is sensitive to $\alpha_s$.

The above results, together with the SM predictions,
are shown in Fig.~\ref{fig:st} as the projection onto
the ($S,\,T$) plane.
Accurate parametrizations of the SM contributions to the
$S$, $T$, $U$ parameters are found in Ref.\citen{hhkm}.
Also shown are the predictions\cite{stu} of
the minimal (one-doublet)
SU($N_c$) Technicolor (TC) models with $N_c\!=\!2,3,4$.
It is clearly seen that the current experiments provide a
fairly stringent constraint on the simple TC models
if a QCD-like spectrum and large $N_c$ scaling
are assumed\cite{stu}.
It is necessary for a realistic TC model to provide an
additional negative contribution to $S$ and a negligibly small
contribution to $T$ at the same time\cite{appelquist93}.

\begin{figure}[t]
\begin{minipage}[t]{7.4cm}
  \leavevmode\psfig{file=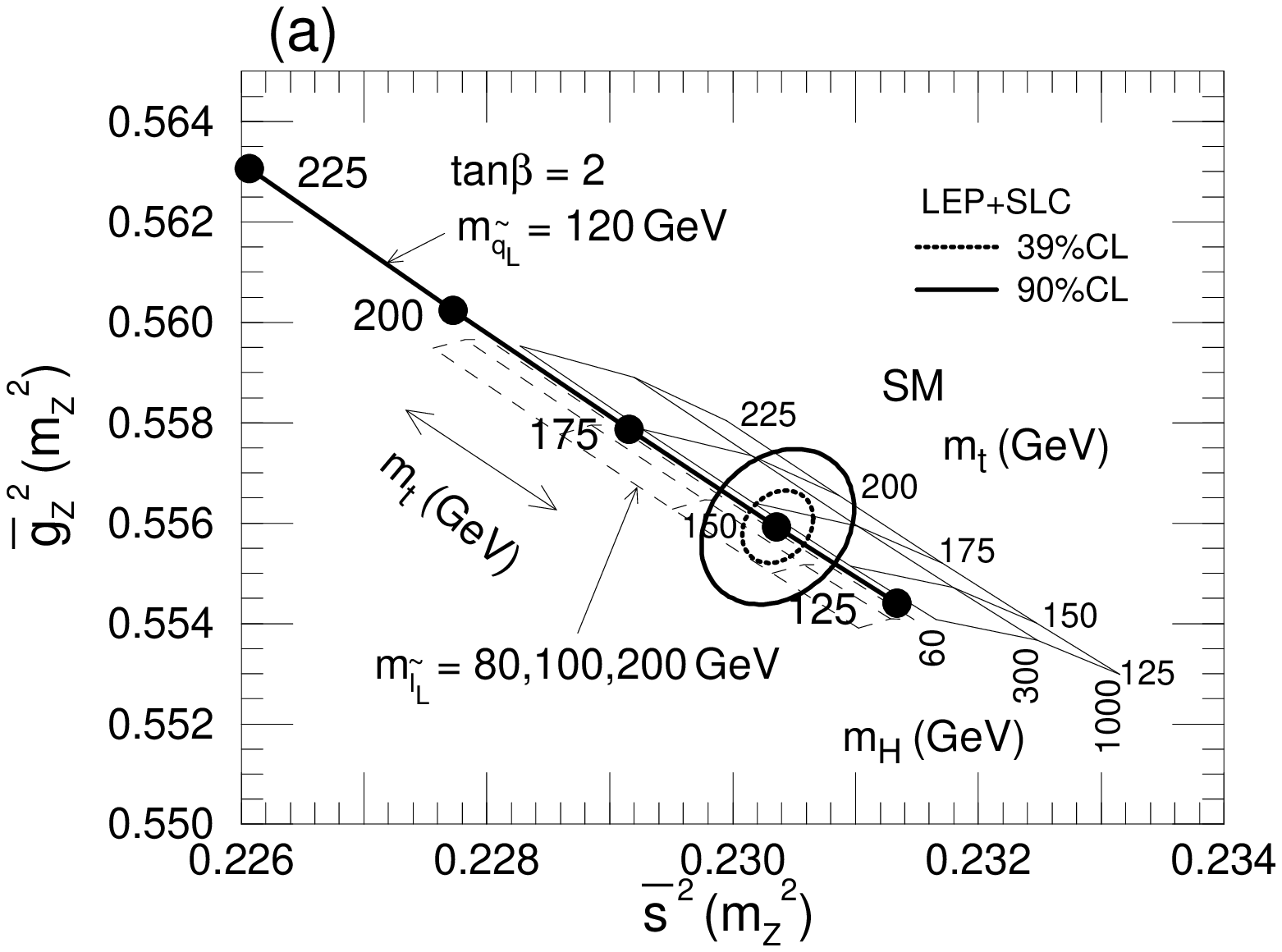,width=7.3cm,silent=0}
\end{minipage}
\hfill
\begin{minipage}[t]{7.4cm}
  \leavevmode\psfig{file=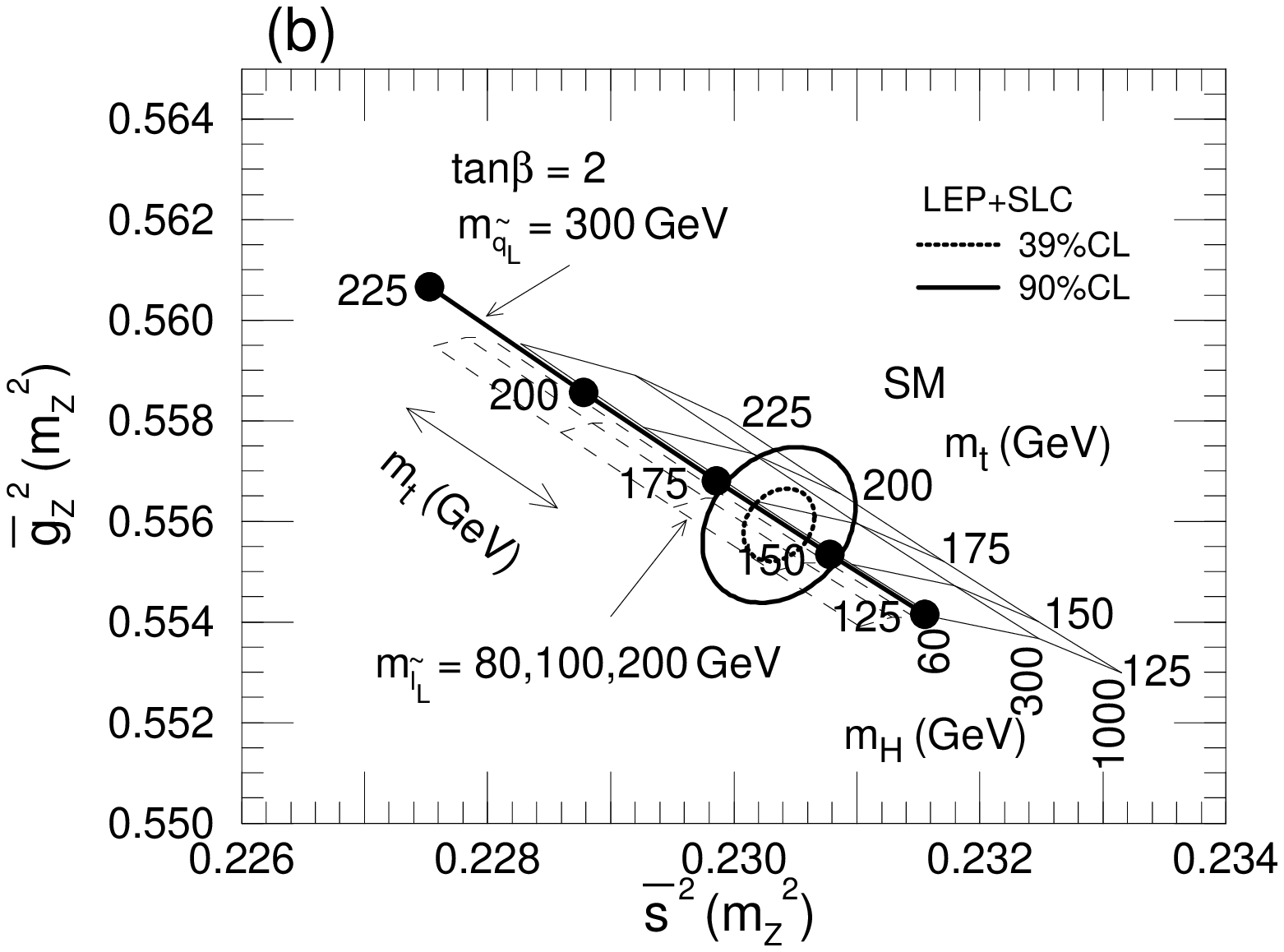,width=7.3cm,silent=0}
\end{minipage}
\caption{%
Minimal SUSY-SM contributions in the
($\sbar^2(\mmz ),\,\gzbar^2(\mmz )$) plane.
The experimental constraint from LEP and SLC is obtained by
assuming $\alpha_s=0.12$ and by using $m_t=175\,$GeV to
calculate the $\zbb$ vertex (the dashed contour in Fig.~2).
}
\label{fig:gzbsb_susy}
\end{figure}

It is interesting to contrast the situation with the
predictions of the minimal SUSY-SM (MSSM).
Shown in Fig.~\ref{fig:gzbsb_susy} are examples of the
MSSM contributions\cite{hmy96} in the
($\sbar^2(\mmz),\,\gzbar^2(\mmz)$) plane.
Contributions of the squarks and sleptons are shown
by thick solid lines and dashed lines, respectively.
It is clearly seen that the SUSY
particle contributions are significant only when
their masses are very near to $\mz$.

Finally, if we regard the point $(S,T,U)=(0,0,0)$ as the
point with no-electroweak corrections, then we find
$\chi^2_{\rm min}/({\rm d.o.f.})=141/(22)$ which has
probability less than $10^{-18}$.
On the other hand, if we also switch-off the remaining
electroweak corrections to $G_F$ by setting $\delg=0$,
then we find $T'=0.0055/\alpha=0.75$, and the point
$(S,T',U)=(0, 0.75, 0)$ gives
$\chi^2_{\rm min}/({\rm d.o.f.})=34.2/(22)$
which is consistent with the data at the 5\%CL.
As emphasized in Ref.\citen{novikov93},
the genuine electroweak correction is not trivial to
establish in this analysis because of the cancellation
between the large $T$ parameter from $m_t \sim 175$~GeV and
the non-universal correction $\delg$ to the muon decay
constant in the observable combination\cite{hhkm}
$T'$ of Eq.(\ref{t'}).

\subsection{ Impact of the Low-Energy Neutral-Current Data }

In this subsection, we show individual contributions from the
four sectors of the low-energy neutral-current data\cite{hhkm},
$\nu_\mu$-$q$ and $\nu_\mu$-$e$ processes, atomic parity
violation (APV), and the classic $e$-$D$ polarization
asymmetry data.

The only new additional data this year is from the CCFR
collaboration\cite{ccfr95_kevin} which measured the ratio of
the neutral-current and charged-current cross-sections
in the $\nu_\mu$ scattering off nuclei.
By using the model-independent parameters of Ref.\citen{fh88},
they constrain the following linear combination,
\begin{subequations} \label{k_ccfr}
\bea
K = 1.732 g_L^2
  + 1.119 g_R^2
  - 0.100 \delta_L^2
  - 0.086 \delta_R^2 \,,
\eea
and find
%
\dataofnqccfrorig
\end{subequations}
The data constrain the charge form factors $\sbar^2(0)$ and $\gzbar^2(0)$.
By combining with all the other neutral-current data of Ref.\citen{hhkm}
we find the fit Eq.(\ref{fitoflenc}).

In order to compare these constraints with those from the
LEP/SLC experiments it is useful to re-express the fit
in the $(\sbar^2(\mmz ),\gzbar^2(\mmz ))$ plane by
assuming the SM running of the charge form-factors.
The combined fit of Eq.(\ref{fitoflenc}) then becomes
%
\fitoflencatmz
%
\vfill
\newpage
\noindent
\begin{wrapfigure}{r}{7.8cm}
 \begin{center}
 \leavevmode\psfig{file=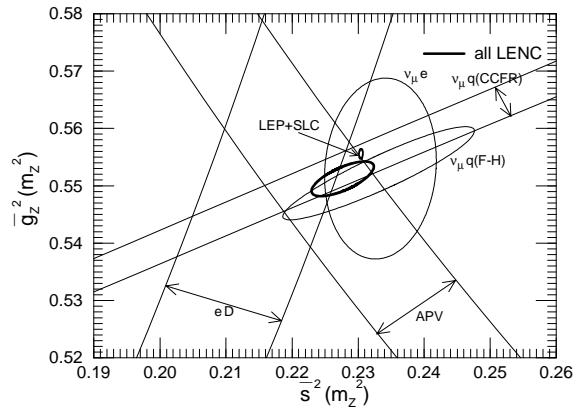,width=7.5cm,silent=0}
 \end{center}
\caption{%
Fit to the low-energy neutral-current data in terms of the
two universal charge form-factors $\protect\sbar^2(\mmz )$ and
$\protect\gzbar^2(\mmz )$.
1-$\sigma$ (39\%CL) contours are shown separately for
the old\protect\cite{fh88}
and the new\protect\cite{ccfr95_kevin}
$\nu_\mu$--$q$ data,
the $\nu_\mu$--$e$ data,
the atomic parity violation (APV) data,
and the SLAC $e$--${\rm D}$
polarization asymmetry data.
The 1-$\sigma$ contour of the combined fit
Eq.(\protect\ref{fitoflencatmz})
is shown by the thick contour.
Also shown is the constraint from the LEP/SLC data,
which is the solid contour in Fig.~\protect\ref{fig:gzbsb}.
}
\label{fig:lenc}
\end{wrapfigure}
%
In Fig.~\ref{fig:lenc} we show individual contributions to
the fit, together with the combined LEP/SLC fit (the solid
contour of Fig.~\ref{fig:gzbsb}).
It is clear that the low-energy data have little impact
on constraining the effective charges, or equivalently
the $S$ and $T$ parameters.
They constrain, however, possible new interactions
beyond the ${\rm SU(2)_L\times U(1)_Y}$ gauge interactions,
such as those from a $Z'$, an additional $Z$ boson.
The mixing-independent constraints on the $Z'$ are obtained e.g.\
from the low-energy neutral-current data\cite{langacker94},
the off-$Z$-resonance neutral-current data from HERA and
TRISTAN\cite{hnst90}, and from its direct production at
the Tevatron\cite{pdg94}.
It is hence highly desirable for all the electroweak experiments
to report model-independent parametrizations of the data,
such as Eq.(\ref{k_ccfr}), rather than to report only
the SM fit results.

\subsection{ The Minimal Standard Model Confronts the Electroweak Data }
In this subsection we assume that all the radiative corrections
are dominated by the SM contributions and obtain constraints
on $m_t$ and $\mh$ from the electroweak data.

\begin{figure}[t]
\begin{minipage}[t]{7.4cm}
  \leavevmode\psfig{file=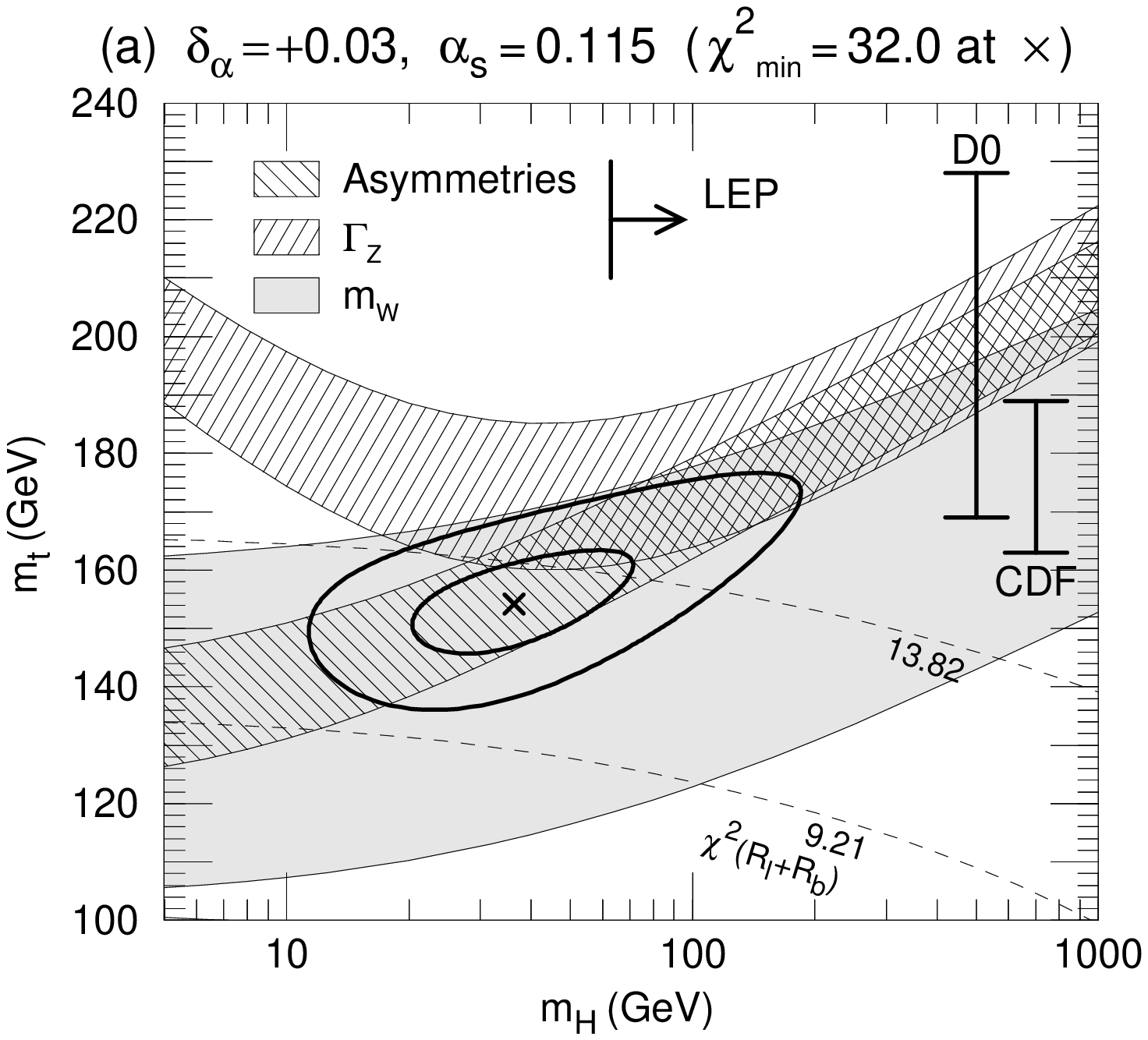,width=7.2cm,silent=0}\\[2mm]
  \leavevmode\psfig{file=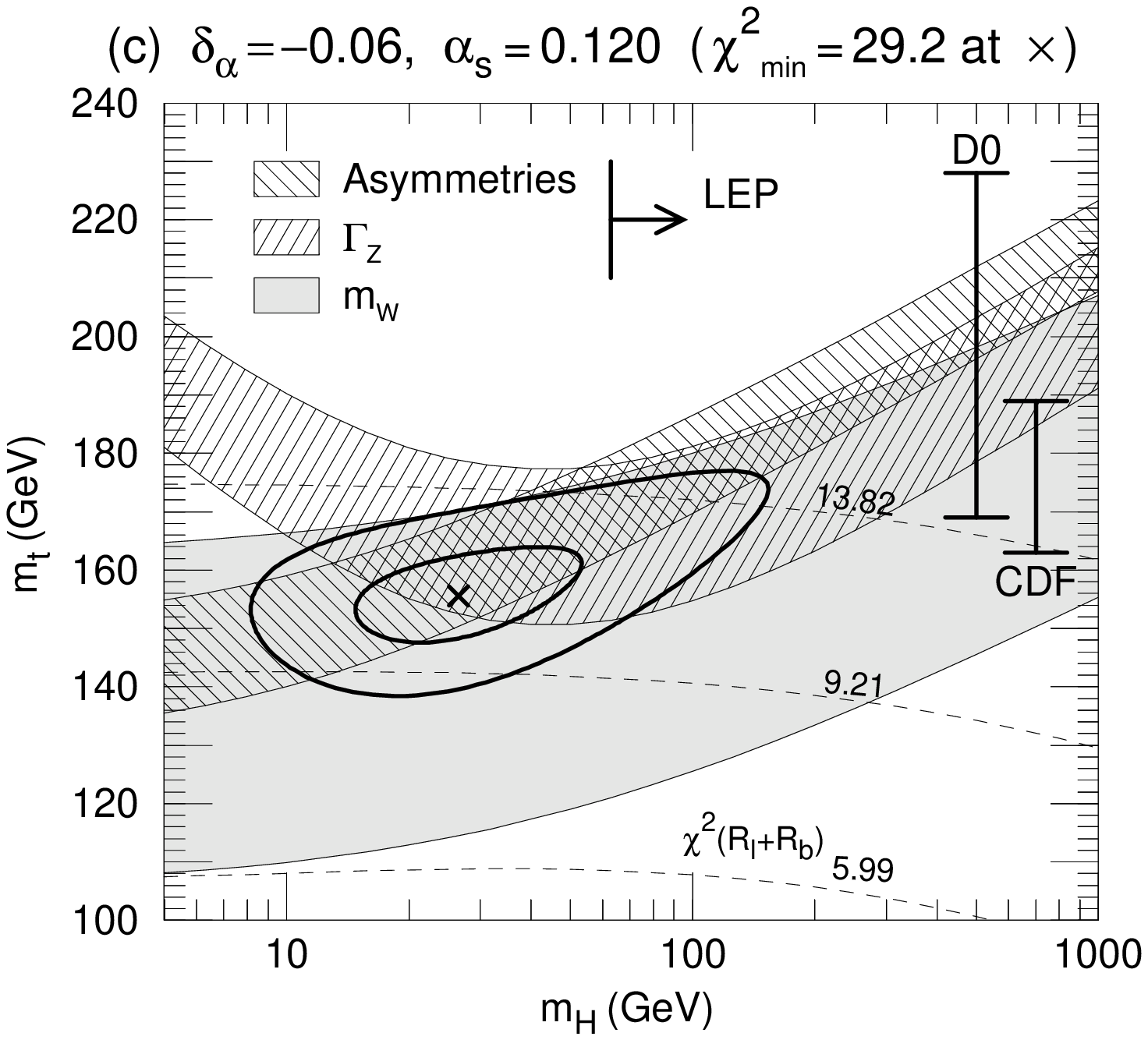,width=7.2cm,silent=0}
\end{minipage}
\hfill
\begin{minipage}[t]{7.4cm}
  \leavevmode\psfig{file=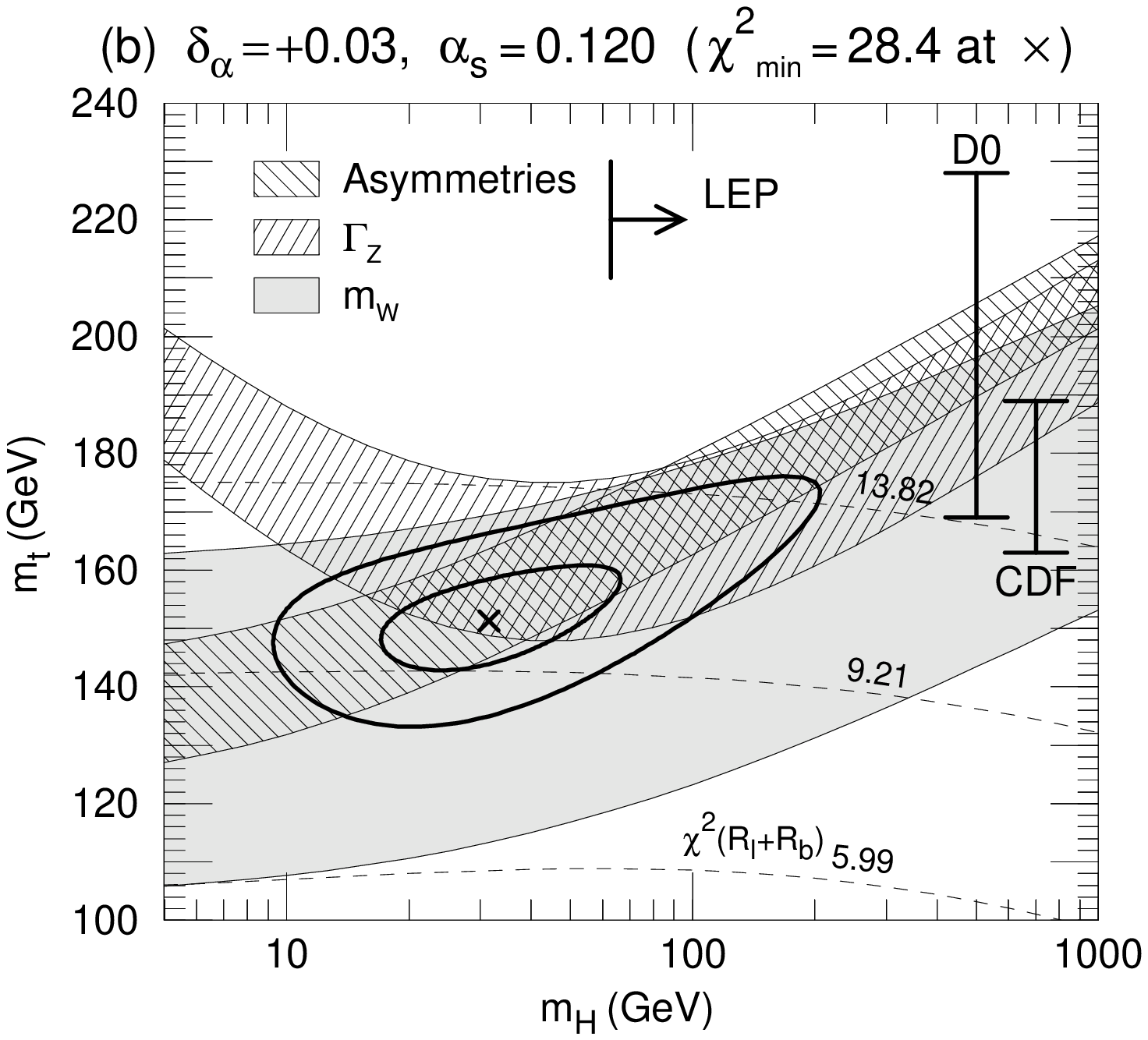,width=7.2cm,silent=0}\\[2mm]
  \leavevmode\psfig{file=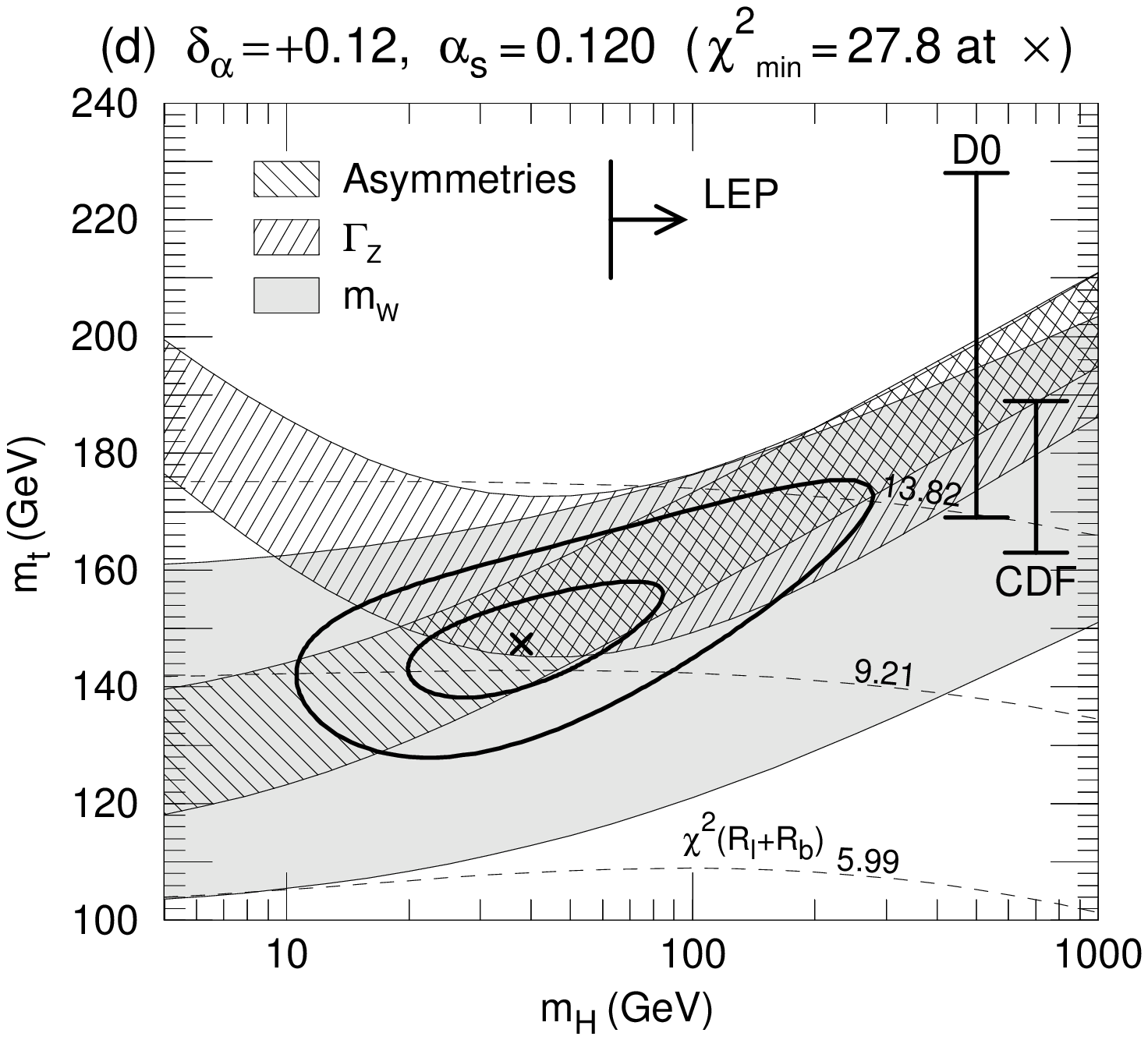,width=7.2cm,silent=0}
\end{minipage}
\caption{%
The SM fit to all electroweak data in the ($\protect\mh,\,m_t$)
plane for
(a) $(\delta_\alpha, \alpha_s)=(+0.03,0.115)$,
(b) $(+0.03,0.120)$,
(c) $(-0.06,0.120)$ and
(d) $(+0.12,0.120)$.
The thick inner and outer contours correspond to
$\Delta\chi^2=1$ ($\sim$ 39\%CL),
and $\Delta\chi^2=4.61$ ($\sim$ 90\%CL), respectively.
The minimum of $\chi^2$ is marked by an ``$\times$'',
where the degree-of-freedom of the fit is 23.
Also shown are the 1-$\sigma$ constraints from the
$Z$-pole asymmetries, $\Gamma_Z$ and $\mw$.
The dashed lines show the constraint from
$R_\ell$ and $R_b$:
$\chi^2\!=5.99,\,9.21,\,13.82$ correspond
to 95\%, 99\%, 99.9\%CL contours, respectively.
}
\label{fig:mtmh}
\end{figure}

In the minimal SM all the form-factors,
$\gzbar^2(\mmz)$, $\sbar^2(\mmz)$, $\gzbar^2(0)$, $\sbar^2(0)$,
$\gwbar^2(0)$ and $\delb(\mmz)$, depend uniquely
on the two mass parameters $m_t$ and $\mh$.
Fig.~\ref{fig:mtmh} shows the result of the global fit to all
electroweak data in the ($\mh,\,m_t$) plane for
(a) $\alpha_s=0.115$ and (b) 0.120 with $\delta_\alpha=0.03$,
and with (c) $\delta_\alpha\!=\!-0.06$
and (d) $+0.12$ for $\alpha_s=0.120$.
The thick inner and outer contours correspond to
$\Delta\chi^2\equiv\chi^2-\chi^2_{\rm min}=1$ (39\%CL),
and $\Delta\chi^2=4.61$ (90\%CL), respectively.
The minimum of $\chi^2$ is
indicated by an ``$\times$'' and the corresponding values
of $\chi^2_{\rm min}$ are given.
We also give the separate 1-$\sigma$ constraints arising from
the $Z$-pole asymmetries, $\Gamma_Z$, and $\mw$.
The asymmetries constrain $m_t$ and $\mh$ through
$\sbar^2(\mmz)$, while $\Gamma_Z$ constrains them through
the three form-factors $\gzbar^2(\mmz)$, $\sbar^2(\mmz)$ and
$\delb(\mmz)$.
In other words, the asymmetries measure the combination of
$S$ and $T$ as in Eq.(\ref{sbar_approx});
both $S$ and $T$ are functions of $m_t$ and $\mh$\cite{hhkm}.
On the other hand, $\Gamma_Z$ measures a different combination
of $S$ and $T$ with an additional contribution from $\delb$.
A remarkable point apparent from Fig.~\ref{fig:mtmh}
is that, in the SM, when $m_t$ and $\mh$ are much larger than
$\mz$, $\Gamma_Z$ depends upon almost the same combination of
$m_t$ and $\mh$ as the one measured through $\sbar^2(\mmz)$.
This is because the quadratic $m_t$-dependence of
$\gzbar^2(\mmz)$ and that of $\delb$ largely cancel in
the SM prediction for $\Gamma_Z$.
Because of this only a band of $m_t$ and $\mh$ can be
strongly constrained from the asymmetries and $\Gamma_Z$ alone
despite their very small experimental errors.
The constraint from the $\mw$ data overlaps this allowed region.

\begin{wrapfigure}{r}{8cm}
\begin{center}
 \leavevmode\psfig{file=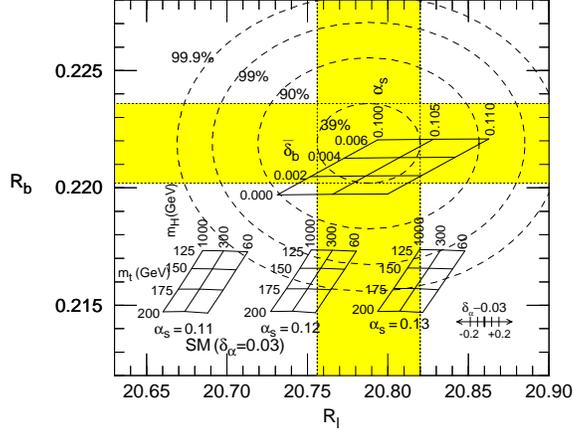,width=7.5cm,silent=0}
\end{center}
\caption{%
Constraints on the $R_b$ vs $R_\ell$ plane.
The SM predictions are shown for
$\alpha_s$=$0.11$, $0.12$, $0.13$
at $\delta_\alpha=0.03$.
The ($\alpha_s,\,\delb$) lattice is obtained
by allowing $\delb$ to vary at
$\sbar^2(\mmz)=0.23039$. 
}
\label{fig:rlrb}
\end{wrapfigure}

Quantities which help to disentangle
the above $m_t$-$\mh$ correlation are $R_\ell$ and $R_b$.
The constraints from these data are shown in
Fig.~\ref{fig:mtmh} by dashed lines corresponding to
$\chi^2=5.99$ (95\%CL),
$\chi^2=9.21$ (99\%CL) and
$\chi^2=13.82$ (99.9\%CL) contours.
These constraints can be clearly seen in Fig.~\ref{fig:rlrb}
where we show the data and the SM predictions for $R_\ell$ and $R_b$.
$R_\ell$ is sensitive to the assumed value of $\alpha_s$, and,
for $\alpha_s=0.120$, the data favors smaller $\mh$.
$R_b$ is, on the other hand, sensitive to neither $\alpha_s$
nor $\mh$, and the data strongly disfavors large $m_t$.
It is thus the $R_\ell$ and $R_b$ data that constrain
the values of $m_t$ and $\mh$ from above.
If it were not for the data on $R_\ell$ and $R_b$
the common shaded region in Fig.~\ref{fig:mtmh}
%
%
with very large $\mh$ $(\mh\sim 1\tev)$ could not
be excluded by the electroweak data alone.
\par
It is clearly seen from Fig.~\ref{fig:mtmh} that
the narrow ``asymmetry''
band is sensitive to $\delta_\alpha$, whereas the ``$\Gamma_Z$''
constraint is sensitive to $\alpha_s$.
The fit improves at larger $\delta_\alpha$ (larger
$1/\bar{\alpha}(\mmz )$) because the ``asymmetry'' constraint
then favors lower $m_t$ that is favored by the $R_b$ data.
An update of the compact parametrization\cite{hhkm} of
the $\chi^2_{\rm SM}$ function of the global fit
has been reported in Ref.\citen{lp95_hag}.
For $\mh=60,300,1000\gev$, $\alpha_s=0.120\pm 0.07$ and
$\delta_\alpha=0.03\pm 0.09$, one obtains
\bea \label{mtfit_standard}
     m_t = 179 \pm 7 {}^{+19(\mh=1000)}_{-22(\mh=60)}
		\mp2 (\alpha_s)
		\mp5 (\delta_\alpha) \,,
\eea
where the mean value is for $\mh=300$~GeV.
The fit (\ref{mtfit_standard}) agrees excellently with
the estimate
\bea \label{mt_tevatron}
	m_t = 180 \pm 13~\gev
\eea
from the direct production data at the Tevatron\cite{mt_cdf,mt_d0}.
Despite the claim\cite{novikov93} that there is no strong
evidence for the genuine electroweak correction, which we
re-confirmed above with the new data,
this should be regarded as strong evidence that the standard
electroweak gauge theory is valid at the quantum level.
The accidental cancellation of the two large radiative effects
in the observable combination $T'$ of Eq.(\ref{t'})
tells us, in the face of the Tevatron results
(\ref{mt_tevatron}), the presence of the large electroweak
correction to the muon-decay rate, $\delg$, which is finite
and calculable only in the gauge theory\cite{sirlin94,lp95_hag}.

\begin{table}[b]
\begin{center}
\caption{ 95\%CL upper and lower bounds of $\mh$(GeV)
for a given $\alpha_s$ and $\delta_\alpha=0.03\pm0.09$.
}
\label{tab:mhlimit}
\vspace*{1mm}

{\small
 \begin{tabular}{|c|clc|clc|clc|}
\hline
$\alpha_s$ &
\multicolumn{3}{|c|}{all EW data} &
\multicolumn{3}{|c|}{$-(R_b,R_c)$ data} &
\multicolumn{3}{|c|}{$+m_t$ (Tevatron)} \\
\hline
0.115 && $16<\mh<150$ &&& $18<\mh<290 $ &&& $22<\mh<360$ &\\ \hline
0.120 && $13<\mh<180$ &&& $15<\mh<500 $ &&& $20<\mh<550$ &\\ \hline
0.125 && $11<\mh<220$ &&& $12<\mh<1800$ &&& $18<\mh<980$ &\\ \hline
\end{tabular}
}

\end{center}
\end{table}


As discussed above the constraint on $\mh$ from the electroweak
data is sensitive to $R_b$, and hence to $\alpha_s$.
Shown in Table~\ref{tab:mhlimit} are the 95\%CL upper and lower
bounds on $\mh$(GeV) from the electroweak data.
A low-mass Higgs boson is clearly favored.
However, this trend disappears for $\alpha_s>0.12$
once we remove the $R_b$ and $R_c$ data.
The present $m_t$ estimate (\ref{mt_tevatron}) from
the Tevatron does not significantly improve the situation.

It is instructive to anticipate the impact a precise
measurement of the top-quark mass would have in the context of
the present electroweak data.
For instance, a precision measurement of $m_t$ with an error of
1~GeV is envisaged\cite{tev33} at TeV33,
\begin{wrapfigure}{r}{8cm}
\begin{center}
  \leavevmode\psfig{file=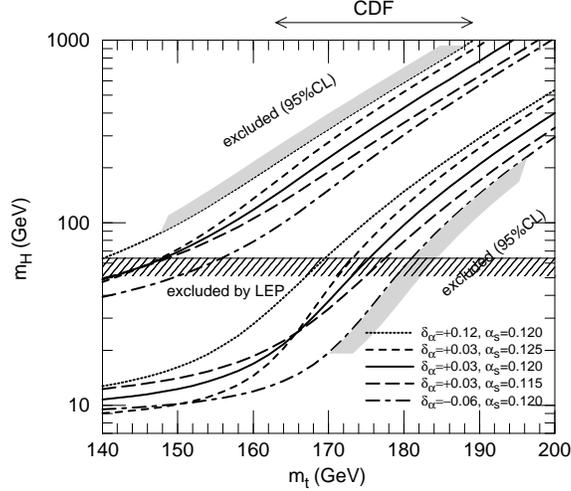,width=7.5cm,silent=0}
\end{center}
\caption{%
Constraints on the Higgs mass in the SM from all the
electroweak data.
Upper and lower bounds of the Higgs mass at 95\%CL are
shown as functions of the top mass $m_t$, where
$m_t$ is treated as an external parameter
with negligible uncertainty.
The results are shown for $\alpha_s=0.120\pm 0.005$ and
$\delta_\alpha\equiv 1/\bar{\alpha}(\mmz )-128.72=0.03\pm 0.09$.
}
\label{fig:mhlimit}
\end{wrapfigure}
a proposed luminosity upgrade of the Tevatron.
The 95\%CL upper/lower bounds on $\mh$ from the electroweak
data are shown in Fig.~\ref{fig:mhlimit} as functions of $m_t$.
Dependences of the bounds on the two remaining parameters,
$\alpha_s=\alpha_s(\mz )_{\msbar}$ and
$\delta_\alpha=1/\bar{\alpha}(\mmz )-128.72$,
are shown clearly.
For a smaller value of $m_t$, $m_t<170$~GeV,
a rather stringent upper bound on $\mh$ is obtained,
whereas a medium-heavy Higgs boson is favored for
$m_t>180$~GeV.
It is tantalizing that the present data from the Tevatron
(\ref{mt_tevatron}) lies just on the boundary.
\par
Fig.~\ref{fig:mhlimit} shows us that once the top-quark mass
is determined, either by direct measurements or by a
theoretical model, the major remaining uncertainty
is in $\delta_\alpha$, the magnitude of the QED
running coupling constant at the $\mz$ scale.
It is clear\cite{lp95_hag} that we won't be
able to learn about $\mh$ in the SM, nor about
physics beyond the SM from its quantum effects,
without a significantly improved determination of
$\bar{\alpha}(\mmz )$.



\section{ The $R_b$ and $R_c$ Crisis and $\alpha_s$ }

\begin{figure}[b]
\begin{minipage}[t]{7.4cm}
  \leavevmode\psfig{file=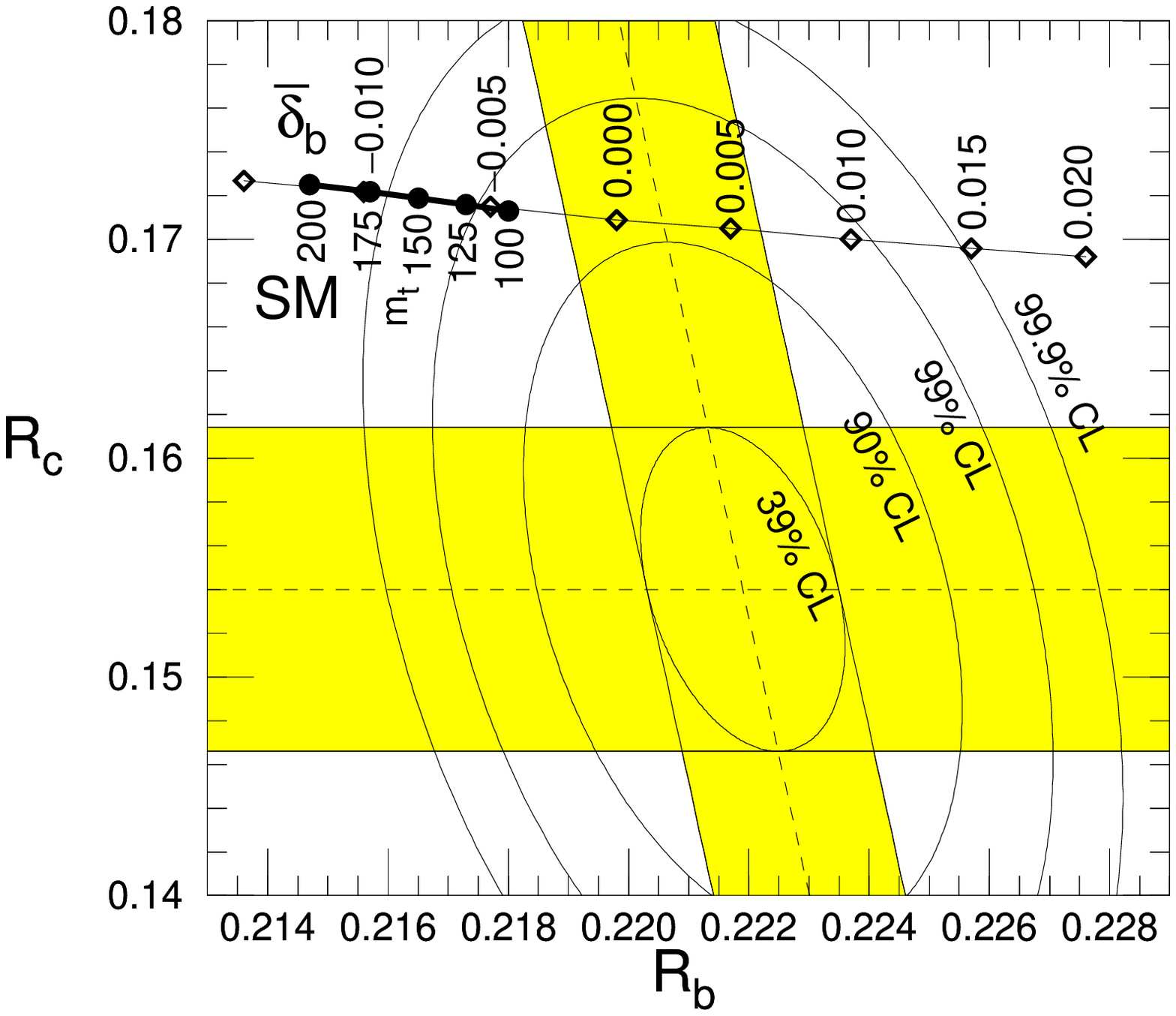,height=6cm,silent=0}
   \vspace{2mm}
\flushright{
  \begin{minipage}[t]{7.1cm}
	\caption{%
	$R_b$ and $R_c$ data 
	and the SM predictions. 
	}
	\label{fig:rbrc}
  \end{minipage}
}
\end{minipage}
\hfill
\begin{minipage}[t]{7.4cm}
  \leavevmode\psfig{file=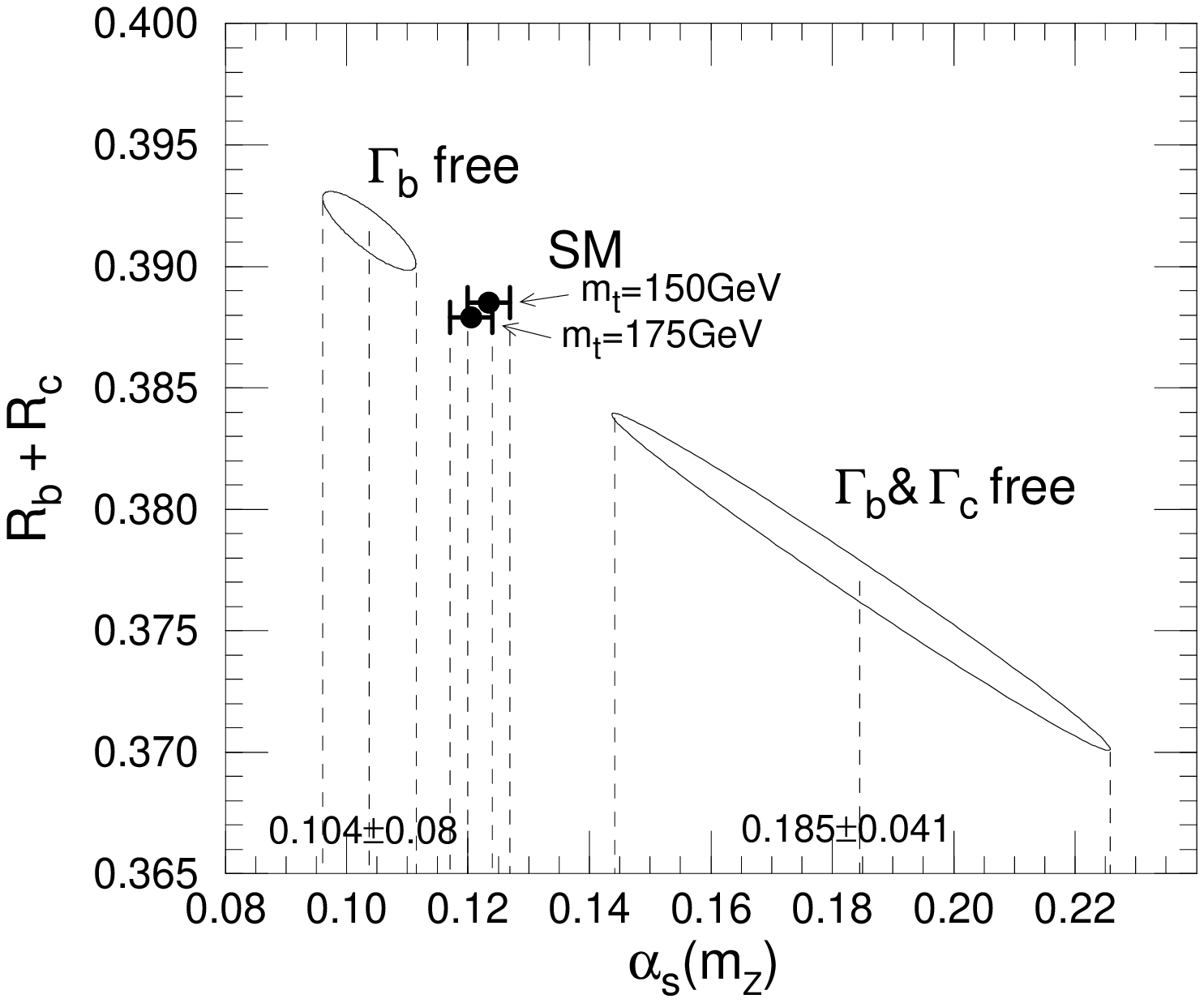,height=6cm,silent=0}
   \vspace{2mm}
\flushright{
  \begin{minipage}[t]{7.1cm}
	\caption{%
	$R_b+R_c$ vs $\alpha_s$.
	}
	\label{fig:rbrc_vs_alphas}
  \end{minipage}
}
\end{minipage}
\end{figure}

The most striking results of the updated electroweak data are
those of $R_b$ and $R_c$, which are shown in Fig.~\ref{fig:rbrc}.
The SM predictions to these ratios are shown by the thick
solid line, where the top-quark mass in the $\zbb$ vertex
correction is indicated by solid blobs.
When combined the $R_b$ and $R_c$ data alone reject the SM
at the 99.99\%CL for $m_t>170$~GeV.
The thin solid line represents the prediction of the extended
SM where the $\zbb$ vertex function, $\delb(\mmz )$ of
(\ref{zblbl}), is allowed to take an arbitrary value.
In the SM the function $\delb(\mmz )$ always takes a negative
value ($\delb(\mmz )\simlt -0.03$), and its magnitude grows
quadratically with $m_t$; see Eq.(\ref{delb_sm}).
The data are not only inconsistent with the SM but also
inconsistent at more than the 2-$\sigma$ level with its
extension where only the $\zbb$ vertex function is modified.

The correlation between the two observables, $R_b$ and $R_c$,
can be understood as follows\cite{lephf9502,lp95_renton}:
To a good approximation, the measurement of $R_c$ does not
depend on the assumed value of $R_b$, because it is measured
by detecting leading charmed-hadrons in a leading-jet for
which a b-quark jet rarely contributes.
On the other hand, the measurement of $R_b$ is affected by
the assumed value of $R_c$, since it typically makes use of
its decay-in-flight vertex signal for which charmed particles
can also contribute.
We find that the following parametrization,
\bsub \label{rb_vs_rc}
\bea
\label{rb_for_rc}
	R_b &=& 0.2205 -0.0136\frac{R_c-0.172}{0.172} \pm0.0016 ,\\
\label{rc_dat}
	R_c &=& 0.1540 \pm 0.0074 ,
\eea
\esub
reproduces the results of Ref.\citen{lephf9502},
as indicated by the shaded regions in Fig.~\ref{fig:rbrc}.

Before discussing the implications of this striking result,
we should recall the fact that the hadronic $Z$ partial width,
$\Gamma_h$, is measured with 0.17\% accuracy;
see Eq.(\ref{del_zwidths}).
This strongly constrains our attempt to modify theoretical
predictions for the ratios $R_b$ and $R_c$.
This is because $\Gamma_h$ can be approximately expressed as
\bea
	\Gamma_h &=& \Gamma_u +\Gamma_d +\Gamma_s +\Gamma_c +\Gamma_b
		+\Gamma_{\rm others}
\nonumber\\
	&\sim& \{ \Gamma_u^0 +\Gamma_d^0 +\Gamma_s^0
	+\Gamma_c^0 +\Gamma_b^0 \}
	\times [1+\frac{\alpha_s}{\pi}+ {\cal O}(\frac{\alpha_s}{\pi})^2 ] ,
\label{gamma_h_appr}
\eea
where the $\Gamma_q^0$'s are the partial widths in the absence of the
final state QCD corrections.
Hence, to a good approximation, the ratios $R_q$ can be expressed
as ratios of $\Gamma_q^0$ and their sum.
A decrease in $R_b$ and an increase in $R_c$ should then
imply a decrease and an increase of $\Gamma_b^0$ and $\Gamma_c^0$,
respectively, from their SM predicted values.
In order to satisfy the experimental constraint on $\Gamma_h$
one should hence adjust the $\alpha_s$ value in Eq.(\ref{gamma_h_appr}).

The consequence of this constraint is clearly shown in
Fig.~\ref{fig:rbrc_vs_alphas} where,
once we allow {\em both} $\Gamma_b^0$ {\em and} $\Gamma_c^0$
to be freely fitted by the data, the above $\Gamma_h$
constraint forces $\alpha_s$ to be unacceptably large,
$\alpha_s(\mz)=0.185\pm0.041$.
On the other hand, if we allow only $\Gamma_b^0$ to vary
by assuming the SM value of $\Gamma_c^0$
(the straight line of the extended SM in Fig.~\ref{fig:rbrc}),
then the $\Gamma_h$ constraint gives a slightly small value,
$\alpha_s(\mz)=0.104\pm0.08$, which is
compatible\cite{hhkm,shifman95}
with some of the low-energy measurements\cite{pdg94,ccfr95}
and lattice QCD estimates\cite{elkhadra92,davies95}.
Although the SM does not reproduce the $R_b$ and $R_c$ data
it gives a moderate $\alpha_s$ value consistent with the
estimates based on the $e^+e^-$ jet-shape
measurements\cite{bethke95,sld95}
and the hadronic $\tau$-decay rate\cite{bethke95}.

In fact we do not yet have a definite clue where
in the region $0.105<\alpha_s(\mz )<0.125$ the true
QCD coupling constant lies.
$\alpha_s\sim 0.12$ is favored from the electroweak
data, if we believe in the SM predictions for $R_b$ and
$R_c$ despite the strong experimental signals.
On the other hand $\alpha_s\sim 0.11$ is favored
if we believe in the SM predictions for $\Gamma_c^0$
while allowing new physics to modify $\Gamma_b^0$.
These two solutions both lead to an acceptable
$\alpha_s$ value at present.
However, once we allow new physics in both $\Gamma_b^0$
and $\Gamma_c^0$ and let them be fitted by the data,
then an unacceptably large $\alpha_s$ follows.

Even if we allow new-physics contributions for both
$\Gamma_b^0$ and $\Gamma_c^0$, we cannot explain
the discrepancies in $R_b$ and $R_c$ because of
the constraint $0.105<\alpha_s(\mz )<0.125$.
The only sensible solution, then, may be to allow
new-physics contributions to all $\Gamma_q^0$
such that their sum stays roughly at the SM value,
e.g.\ by making all the down-type-quark widths larger
than their SM values by 3\% and the up-type-quark
widths to be smaller by 6\%.
Such a model would explain both $R_b$ and $R_c$,
and give a reasonable $\alpha_s$.
It is not easy to find a working model, however, which does
not jeopardize all the excellent successes of the SM
in the quark and lepton asymmetries, the leptonic widths,
$\mw$, and in the low-energy neutral-current data.

\begin{figure}[b]
\begin{minipage}[t]{7.4cm}
  \leavevmode\psfig{file=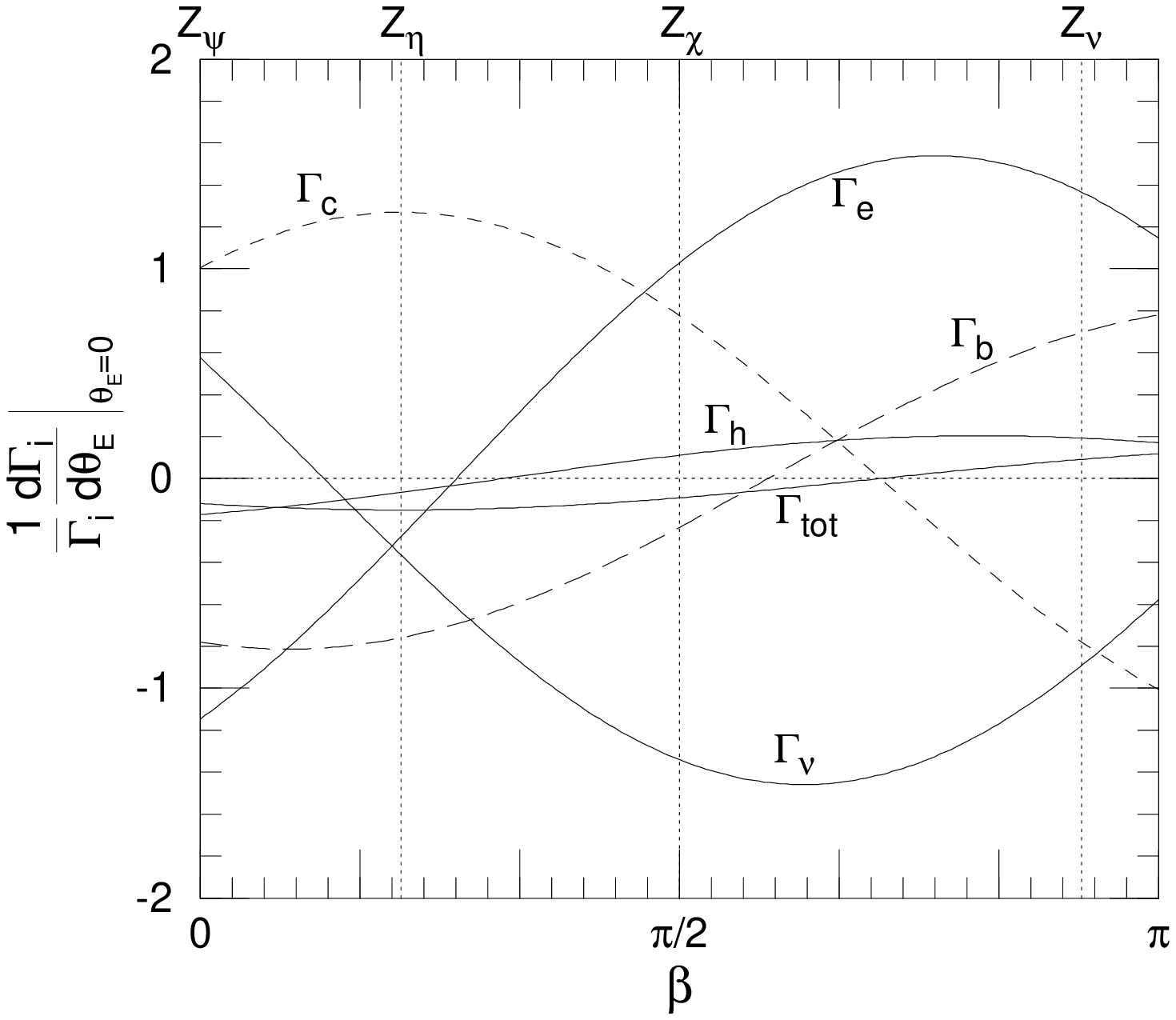,width=7.1cm,silent=0}
\flushright{
  \begin{minipage}[t]{7.1cm}
	\caption{%
	The logarithmic derivative of the partial and total
	widths with respect to the mixing angle versus $\beta$.
	}
	\label{fig:width_vs_beta}
  \end{minipage}
}
\end{minipage}
\hfill
\begin{minipage}[t]{7.4cm}
  \leavevmode\psfig{file=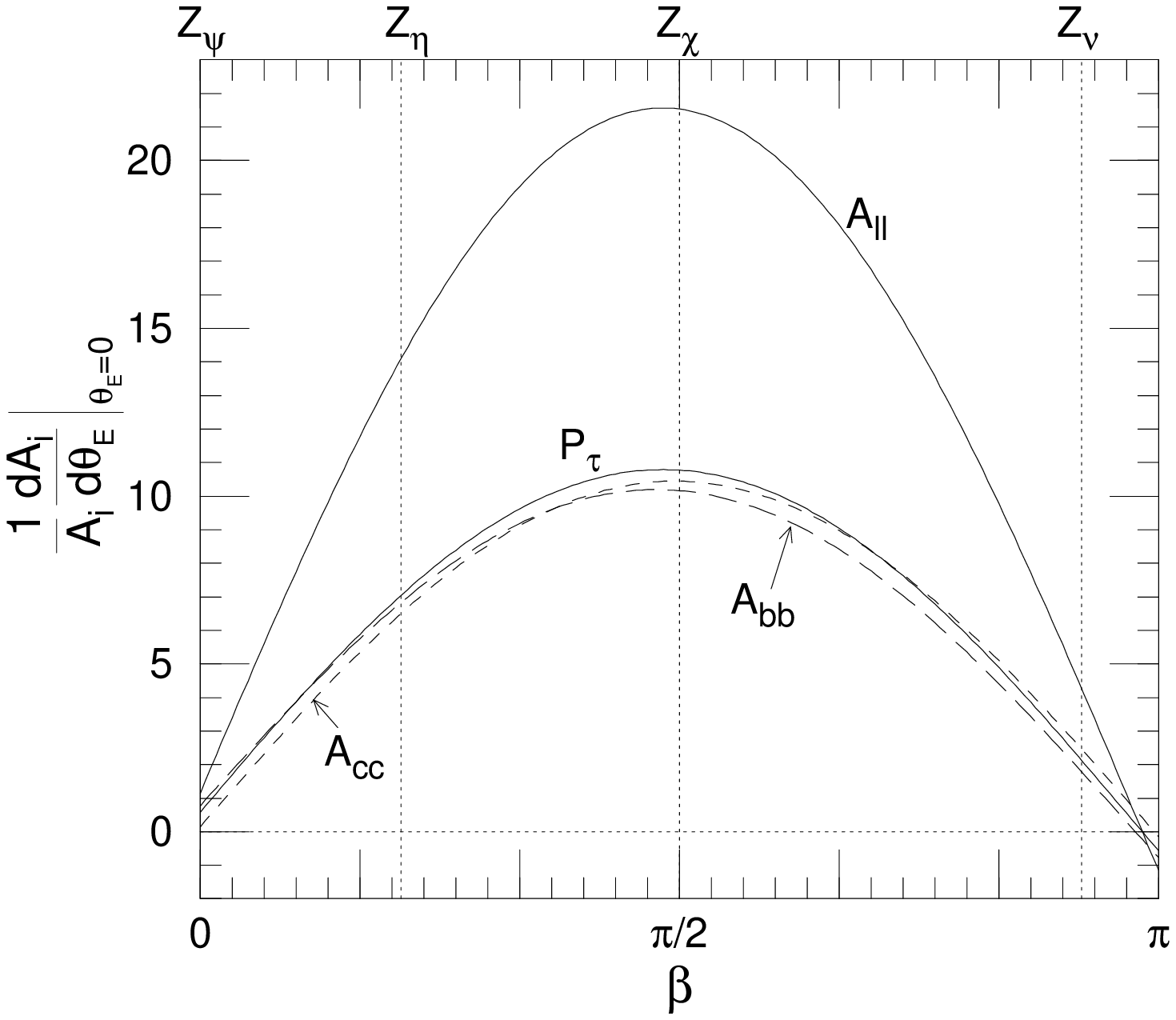,width=7.1cm,silent=0}
\flushright{
  \begin{minipage}[t]{7.1cm}
	\caption{%
	The logarithmic derivative of the asymmetries
	widths with respect to the mixing angle versus $\beta$.
	}
	\label{fig:asym_vs_beta}
  \end{minipage}
}
\end{minipage}
\end{figure}

The difficulty is exemplified in Figs.~\ref{fig:width_vs_beta} and
\ref{fig:asym_vs_beta}, where the effects of the $Z$-$Z'$ mixing
in the partial $Z$ widths and all the on-peak asymmetries are given
for the models with an additional $Z$ boson in the ${\rm E_6}$
grand-unified theory.
The additional $Z$ boson, $Z_E$, is a linear combination of
the SU(5) singlet, $Z_\chi$, and the SO(10) singlet, $Z_\psi$, as
\bea \label{z_e}
Z_E = Z_\psi \cos\beta + Z_\chi \sin\beta \,,
\eea
and the observed $Z$ boson, $Z_1$, is a mixture of the SM $Z$
boson, $Z_{\rm SM}$, and $Z_E$;
\bea \label{z_1}
Z_1 = Z_{\rm SM} \cos\theta_E + Z_E \sin\theta_E \,,
\eea
in the notation of Ref.\citen{hnst90}.
The models with $\beta \approx \pi$ can, for instance, make $R_b$
larger by 4\% and $R_c$ smaller by 5\%, while roughly keeping
$\Gamma_h^0$, $\Gamma_Z^0$ and the on-peak asymmetries.
However, these models should necessarily predict unacceptably
large $\Gamma_\ell$ (by $\sim 6\%$) and too small
$\Gamma_{\rm inv}$ (by $\sim 3\%$), in conflict with the
severe constraints (\ref{del_zwidths}).

Because the $R_c$ measurement depends strongly on the
charm-quark detection efficiency,\, which has uncertainties
in the charmed-quark fragmentation function into charmed
hadrons and in charmed-hadron decay branching fractions,
it is still possible that unexpected errors are hiding.
As an extreme example, if as much as 10\% of the charmed-quark
final states were unaccounted for in the simulation program,
then {\em both} the 10\% deficit in $R_c$ at LEP {\em and}
the 10\% too few charmed hadron multiplicity in B-meson
decays\cite{buchalla95,neubert95} could be solved.

If we assume that $R_c$ actually has the SM value $R_c\sim 0.172$
and temporarily set aside its experimental constraint,
then the correlation as depicted by Eq.(\ref{rb_for_rc}) tells
that the measured $R_b=0.2205\pm0.0016$ is about 2\% larger
than the SM prediction, $R_b\sim 0.216$ for $m_t\sim 175$~GeV.
The discrepancy is still significant at the 3-$\sigma$ level.

I looked for the possibility that an experimental problem
causing an underestimation of $R_c$ could lead to
an overestimation of $R_b$.
This turned out to be a difficult task, since $R_b$ is measured
mainly by using a different technique, the double-tagging
method\cite{lephf9502,lp95_renton}, where the $b$-quark tagging efficiency is
determined experimentally rather than by estimating it from
the $b$-quark fragmentation model and the $b$-flavored
hadron decay rates.  Schematically the single and double
$b$-tag event rates ($N_T$ and $N_{TT}$, respectively)
in hadronic two-jet events are expressed as
\bsub \label{b_tagg}
\bea
\frac{N_T}{2N_h} &=&  	 \epsilon_b r_b R_b
			+\epsilon_c r_c R_c
			+\mbox{\rm others} ,
\\
\frac{N_{TT}}{N_h} &=&  C\,\{\,\epsilon_b^2 r_b R_b
			+\epsilon_c^2 r_c R_c
			+\mbox{\rm others} \,\} ,
\eea
\esub
where $\epsilon_q$ denotes the efficiency of tagging a $q$-jet,
$r_q$ is the rate of two-jet-like events ($\mbox{\it Thrust}>0.8$) in the
$q\bar{q}$ initiated events, and the deviation of $C$ from unity
measures possible correlation effects between the two jets.
By choosing the tagging condition such that
$\epsilon_b \gg \epsilon_c$ one can self-consistently
determine both $\epsilon_b$ and $R_b$:
\bea
\label{rb_btagg}
R_b\, = \,\frac{C}{r_b}\,
\frac{[\,\frac{N_T}{2N_h} -\epsilon_c\frac{r_c}{r_b}R_c -\cdot\cdot\cdot\,]^2}
   {[\,\frac{N_{TT}}{N_h} -\epsilon_c^2\frac{r_c}{r_b}R_c -\cdot\cdot\cdot\,]}
\,.
\eea
In the limit of uncorrelated two-jet events only ($C=1$ and $r_b=1$),
and in the limit of a negligible contribution from non-$b$ events
($\epsilon_c/\epsilon_b=0$),
the ratio $R_b$ is determined from the ratio of the square of
the single-tag event rate and the double-tag event rate.
Only for the corrections to this limit are the QCD motivated
hadron-jet simulation programs used.
A compilation of very careful tests of these correction terms are
found in the LEP/SLC Heavy Flavor Group report\cite{lephf9502}.
We should still examine if our present understanding of generating
hadronic final states from quark-gluon states allows us to constrain
the coefficients $C$ and $r_b$ at much less than one \% level
and the miss-tagging efficiency $\epsilon_c\sim 0.01$ at the 10\% level.
For instance, the combination of an overestimation of $C$ by 0.5\%
with an underestimation of $r_b$ by 0.5\% and that of $\epsilon_c$ by 10\%,
can result in an overestimate of $R_b$ by 2\%.
Serious theoretical studies of the uncertainty in the present
hadron-jet generation program are needed, because in my opinion,
these programs have never been tested at the accuracy level that
was achieved by these excellent experiments at LEP.


\vfill
\newpage

\section{ Attempts to Explain Large $R_b$ }
%
There have been many attempts to explain the discrepancy in
$R_b$ by invoking new physics beyond the SM.
Most notably, in the minimal supersymmetric (SUSY) SM\cite{boul,abc,%
wells1,kimpark,sola1,rc,dabel,wells2,chankow2,wagner,wang,ma,wells3,yhm95},
an additional loop of a light $\tilde{t}_R$ and a light
higgsino-like chargino, or that with an additional Higgs pseudoscalar
when $\tan\beta\gg 1$, can compensate the large negative
top-quark contribution of the SM in the $Zb_Lb_L$ vertex function.
Such a solution typically leads to the prediction that the masses
of the lighter $\tilde{t}$ and chargino, or the pseudoscalar
should be smaller than $\mz$.
In the former scenario the top quark should have significant exotic
decays into $\tilde{t}_R$ and a neutral Higgsino, and in the latter
scenario another exotic decay, $t\to b+H^+$, may
occur\cite{wells2,wang,ma,wells3,yhm95}.
In both SUSY scenarios we should expect to find new particles at
the Tevatron, LEP2 or even at LEP1.5.

It is worth remarking here that
the small $\alpha_s$ value which is obtained by allowing
\begin{wrapfigure}{r}{7cm}
\begin{center}
  \leavevmode\psfig{file=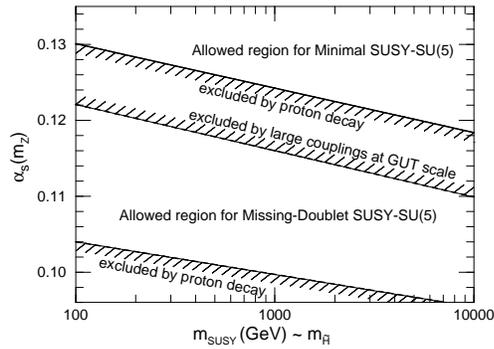,width=6.5cm,silent=0}
\caption{%
Constraints on $\alpha_s(\mz)_{\msbar}$ as functions of the
SUSY threshold scale $m_{\rm SUSY}$ in the minimal SUSY-SU(5)
model 
and in the model with the missing-doublet
mechanism. 
}
\label{fig:gut_vs_alps}
\end{center}
\end{wrapfigure}
a new physics contribution to explain
the $R_b$ anomaly tends to destroy the SUSY-SU(5) unification
of the three gauge couplings in the minimal model\cite{minsu5}.
This problem is, however, highly dependent on details of
the particle mass spectrum at the GUT scale.
In fact in the missing-doublet\cite{mdmsu5} SUSY-SU(5) model
which naturally explains the doublet-triplet splitting,
smaller $\alpha_s$ is prefered due to its peculiar
GUT particle spectrum\cite{hy92,bagger95,lopez95}.
I show in Fig.~\ref{fig:gut_vs_alps}
the update for the allowed regions
of $\alpha_s(\mz )$ in the two SUSY-SU(5) models as functions
of the heavy Higgsino mass, where the standard supergravity model
assumptions are made for the SUSY particle masses at the
electroweak scale\cite{hmy96}.

\begin{figure}[t]
\begin{center}
 \leavevmode\psfig{file=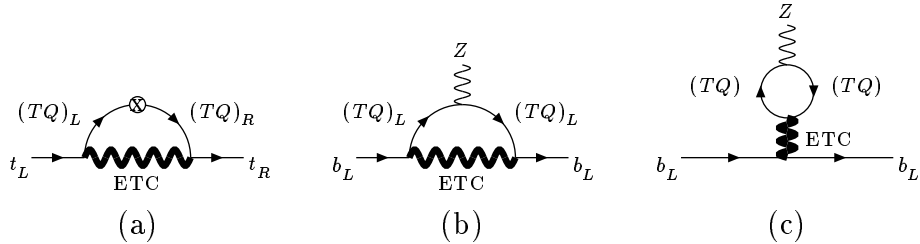,silent=0}
\vspace{2mm}
\caption{%
The side-ways ETC (Extended Techni-Color) bosons
contribute to both (a) the top-quark mass and
(b) the $\zbb$ vertex.
(c) Diagonal ETC bosons can also contribute
to the $\zbb$ vertex.
}
\label{fig:etc}
\end{center}
\end{figure}

In an alternative scenario of electroweak symmetry breaking,
the Techni-Color (TC) model, the heavy top-quark mass implies
strong interactions among top-quarks and techniquarks.
Such interactions, typically called the extended technicolor
(ETC) interactions, can affect the $Zb_Lb_L$ vertex:
see Fig.~\ref{fig:etc}.
However, the side-ways ETC bosons that connect the top-quark and
techniquark leads to a contribution with an opposite
sign\cite{chivukula92,evans94}.
The diagonal ETC bosons contribute\cite{kitazawa93} with
the correct sign\cite{ghwu95}, and their phenomenological
consequences have been studied\cite{yue95,hk95}.
Although the diagonal ETC bosons can explain the $R_b$
data\cite{hk95},
it has been pointed out\cite{yoshikawa95} that such models should
necessarily give an unacceptably large contribution to either the
$S$ or the $T$ parameter.
See Ref.\citen{kitazawa95} for more discussions on these models.

As an alternative to the standard ETC model where the ETC
gauge group commutes with the SM gauge group, models with
a non-commuting ETC gauge group have been proposed\cite{chivukula94}
which have rich phenomenological consequences\cite{simmons95}.

So far, the above models affect mainly the $\zbb$ coupling,
which dominates the $Zb_R^{}b_R^{}$ coupling in the SM.
A possible anomaly in the $b$-jet asymmetry parameter, $A_b$,
observed at the SLC with its polarized beam
(see Fig.~\ref{fig:sb}) may suggest a new physics contribution
in the $Zb_R^{}b_R^{}$ vertex\cite{peskin95}.
It is worth watching improved $A_b$ measurements at the SLC
in the near future.

It has also been proposed\cite{holdom94} that a new heavy gauge
boson $X$ which couples only to the third-generation quarks and
leptons may affect the $Z$ boson experiments through its mixing
with the SM $Z$.
Such models affect not only the $Zbb$ vertex but also
the $Z\tau\tau$ and $Z\nu_\tau \nu_\tau$ vertices.
The original proposal\cite{holdom94} with the axial-vector $Xbb$
coupling and the vector $X\tau\tau$ coupling should be re-examined
in view of the $e$-$\mu$-$\tau$ universality of the 1995 data,
Eq.(\ref{e_mu_tau}), and the possible anomaly in $A_b$.


\def\su{\ifmmode{\tilde{u}} \else{$\tilde{u}$} \fi}
\def\sd{\ifmmode{\tilde{d}} \else{$\tilde{d}$} \fi}
\def\sq{\ifmmode{\tilde{q}} \else{$\tilde{q}$} \fi}
\def\sg{\ifmmode{\tilde{g}} \else{$\tilde{g}$} \fi}
\def\snu{\ifmmode{\tilde{\nu}} \else{$\tilde{\nu}$} \fi}
\def\se{\ifmmode{\tilde{e}} \else{$\tilde{e}$} \fi}
\def\smu{\ifmmode{\tilde{\mu}} \else{$\tilde{\mu}$} \fi}
\def\sfp{\ifmmode{\tilde{f}_{1L}} \else{$\tilde{f}_{1L}$} \fi}
\def\sfm{\ifmmode{\tilde{f}_{2L}} \else{$\tilde{f}_{2L}$} \fi}

\section{Quark-Lepton Universality Violation in Charged Currents}

The tree-level universality of the charged-current weak
interactions is one
of the important consequences of the ${\rm SU(2)_L}$ gauge symmetry.
The universality between quark- and lepton-couplings is expressed
as the unitarity of the Cabibbo-Kobayashi-Maskawa (CKM) matrix.
After correcting for the SM radiative corrections,
the present experimental data\cite{towner94,sirlin95} gives
\bea
\label{unitarity}
|V_{ud}|^2+|V_{us}|^2+|V_{ub}|^2-1 = - 0.0017 \pm 0.0015\,.
\eea
%
\begin{wrapfigure}{r}{7.3cm}
\begin{center}
  \leavevmode\psfig{file=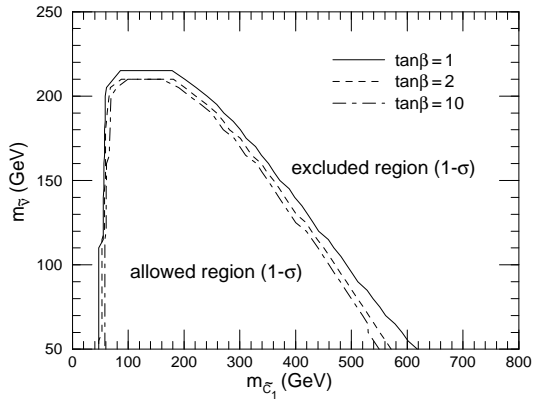,width=7cm,silent=0}
\caption{%
The 1-$\sigma$ allowed region in the
($m_{\tilde{C}_1}$, $m_{\tilde{\nu}}$)
plane from the universality violation
(\protect\ref{unitarity}).
}
\label{fig:susy_cc}
\end{center}
\end{wrapfigure}
Universality is violated at the 1-$\sigma$ level.
\par
Here again, the most important observation is that the quark-lepton
universality of the $W$ couplings is verified experimentally at the
0.2\% level.
This excellent agreement is found only after the SM radiative
corrections are applied\cite{sirlin95}.
Eq.(\ref{unitarity}) therefore strongly constrains any new
interactions
which distinguish between quarks and leptons.
\par
It turned out that the 0.1\% level of the non-universality is
expected in the MSSM, where the expected mass differences
between squarks and sleptons can naturally lead to the
quark-lepton non-universality
at the one-loop level\cite{barbieri85,barbieri92,hmy95}.
The effect, however, disappears quickly if the relevant SUSY particle
masses are much bigger than $\mz$.
\par
Shown in Fig.~\ref{fig:susy_cc} is the 1-$\sigma$ allowed
region of masses of the sneutrino \snu and the lighter
chargino, $\tilde{C}_1$, in the MSSM.
The ratio of the two vacuum expectation values is set at
$\tan\beta=(1,2,10)$.
The 1-$\sigma$ upper bounds are roughly
$m_{\snu}<220$~GeV and $m_{\tilde{C}_1}<600$~GeV, respectively.
It has been found\cite{hmy95} that the sign and the magnitude
of the quark-lepton universality violation (\ref{unitarity})
favor light sleptons and relatively light chargino and neutralinos
with significant gaugino components.
Details are found in Refs.\citen{hmy95,yhm95}.



\section{Conclusions}

The most important hint of physics beyond the Standard Model (SM)
from the 1995 precision electroweak data is that the most precisely
measured quantities, the total, leptonic and hadronic decay widths
of the $Z$ and the effective weak mixing angle, $\sin^2\theta_W$,
measured at LEP and SLC, and the quark-lepton universality of
the weak charged currents measured at low energies, all agree
with the predictions of the SM at a few $\times 10^{-3}$ level.
Any attempts to replace the SM by a more fundamental theory
should hence explain why quantum effects from new physics
are tiny for these quantities.

A most natural explanation seems to me that the new physics
beyond the SM is essentially weakly interacting, and
its quantum effects decouple.
The supersymmetric (SUSY) SM is a good example of such a model.

On the other hand, if new physics has a strongly interacting
sector one may need a mechanism that naturally hides its
potentially large quantum effects.
The technicolor (TC) scenario of the dynamical electroweak
symmetry breaking belongs to this class, but I do not know
of a mechanism that naturally kills all additional quantum
corrections.

In this report I examined implications of the three possible
discrepancies between experiments and the SM predictions.
They are the 11\% (2.5-$\sigma$) deficit of the
$Z$-partial-width ratio $R_c=\Gamma_c/\Gamma_h$,
the 2\% (3-$\sigma$) excess of the ratio
$R_b=\Gamma_b/\Gamma_h$,
and the 0.17\% (1-$\sigma$) deficit of the CKM unitarity
relation $|V_{ud}|^2+|V_{us}|^2+|V_{ub}|^2=1$.

I first explained in detail the difficulty in interpreting
the present $R_c$ data.
Because the $Z$ hadronic width, $\Gamma_h$, agrees with the
SM prediction at the 0.2\% level, the 11\% deficit in $R_c$
should necessarily lead to an unacceptably large $\alpha_s$
if the other hadronic widths of the $Z$ boson remain unaffected.
An acceptable value of $\alpha_s$ is obtained only if the deficit
in $\Gamma_c$ is compensated rather accurately by an excess in
the rest of the hadronic widths
$\Gamma_u+\Gamma_d+\Gamma_s+\Gamma_b$.
In addition, we must reproduce all the quark and lepton
asymmetry data, the leptonic widths and the other successful
predictions of the SM.
I was unable to find such a solution.

If we assume the SM value for $R_c$, then we should identify
the cause of an over-estimation in its experimental determination.
It is possible that such an over-estimation occurs through
several systematic errors in the input parameters that determine
the detection efficiency of the primary charm-quark events
commonly adopted by all experiments at LEP.
The $R_b$ data may, nevertheless, be taken seriously,
since they are obtained by using the double tagging technique
that allows us to measure the $b$-quark-jet tagging efficiency.
It is not clear to me, however, if our understanding of
hadron-jet formation is good enough to establish the 2\%
excess in $R_b$ at the 3-$\sigma$ level.

If we take the present $R_b$ data, it gives us a 3-$\sigma$
evidence for physics beyond the SM.
It may either signal a relatively large additional quantum
correction or a new tree-level neutral-gauge-boson interaction
that couples to $b$.
In the former case, the large radiative effect may either
come from the large Yukawa coupling in the SUSY-SM
or the new strong interaction that makes the top-quark massive
in the Technicolor (TC) scenario of the electroweak gauge
symmetry breaking.
In the SUSY-SM, one is forced to have either a combination of
a very light $\tilde{t}_R$ and Higgsino or that of the large
$b$-quark Yukawa coupling and a very light Higgs boson.
In either case, one expects to observe new particles at
LEP2 or in the top-quark decay at the Tevatron.
It is worth noting that these SUSY scenarios do not cause
any difficulty in the other sector of precision electroweak
physics.
In the TC scenario, one typically obtains large radiative
effects to the $Zb_Lb_L$ vertex from the ETC-gauge-boson
interactions that give rise to the top-quark mass.
The side-ways-boson exchange contributes with the wrong
sign, while the diagonal-boson exchange contributes with
the correct sign.
The naive ETC models, however, suffer from either too
large an $S$ for models with custodial SU(2) symmetry or
too large a $T$ for models without custodial SU(2).
Models with the ETC gauge group that contains the
electroweak ${\rm SU(2)_L}$ might evade such immediate
difficulty.

Finally, I examined consequences of the possible violation
of quark-lepton universality in low-energy
charged-current interactions.
Although the observed deficit in the CKM unitarity relation
is only 0.17\% at the 1-$\sigma$ level, we showed
that this is what the minimal SUSY-SM predicts if
sleptons and gauginos are sufficiently light.

In conclusion, none of the above three hints of deviations
from the SM predictions are convincing to me.
The first two data, $R_c$ and $R_b$, are experimentally
significant but they do not lead us to a satisfactory
picture of physics beyond the SM, yet.
The last example, the SUSY-SM explanation of the violation
of the CKM unitarity relation, can naturally co-exist with
all the other excellent successes of the SM, but its
experimental significance is marginal.

I feel strongly that the experimental accuracy of the precision
electroweak measurements has finally reached the level where
new physics can be probed via its quantum effects.
The $R_b$ data, if even more firmly established, would pose
a serious challenge to theorists.
Intimate and attentive cooperation between experimentalists
and theorists is needed more than ever.

\section*{Acknowledgements}
I would like to thank
D.~Haidt, J.~Kanzaki, N.~Kitazawa, S.~Matsumoto,
T.~Mori, M.~Morii, R.~Szalapski and Y.~Yamada
for fruitful collaborations which made this presentation possible.
I would also like to thank
D.~Charlton, M.~Drees, R.~Jones,
C.~Mariotti, A.D.~Martin, K.~McFarland,
D.R.O.~Morrison, H.~Murayama, B.~Pietrzyk, P.B.~Renton,
M.H.~Shaevitz, D.~Schaile, M.~Swartz,
T.~Takeuchi, P.~Vogel, P.~Wells and D.~Zeppenfeld
for discussions that helped me understand the experimental data
and their theoretical implications better.


\newcommand {\AP}[3]	{ Ann.\ Phys.\ {\bf #1} (#2) #3}
\renewcommand {\CMP}[3]	{ Comm.\ Math.\ Phys.\ {\bf #1} (#2) #3}
\newcommand {\IBID}[3]	{{\it ibid.} {\bf #1} (#2) #3}
\newcommand {\MPL}[3]	{ Mod.\ Phys.\ Lett.\ {\bf #1} (#2) #3}
\renewcommand {\NC}[3]	{ Nuovo Cimento {\bf #1} (#2) #3}
\newcommand {\NPB}[3]	{ Nucl.\ Phys.\ {\bf B#1} (#2) #3}
\renewcommand {\PL}[3]	{ Phys.\ Lett.\ {\bf #1} (#2) #3}
\newcommand {\PLB}[3]	{ Phys.\ Lett.\ {\bf B#1} (#2) #3}
\renewcommand {\PR}[3]	{ Phys.\ Rev.\ {\bf #1} (#2) #3}
\newcommand {\PRD}[3]	{ Phys.\ Rev.\ {\bf D#1} (#2) #3}
\renewcommand {\PRL}[3]	{ Phys.\ Rev.\ Lett.\ {\bf#1} (#2) #3}
\renewcommand {\PTP}[3]	{ Prog.\ Theor.\ Phys.\ {\bf#1} (#2) #3}
\newcommand {\RMP}[3]	{ Rev.\ Mod.\ Phys.\ {\bf#1} (#2) #3}
\newcommand {\ZP}[3]	{ Z.~Phys.\ {\bf #1} (#2) #3}
\newcommand {\ZPC}[3]	{ Z.~Phys.\ {\bf C#1} (#2) #3}

\vfil

\end{document}